\documentclass[iop,apj]{emulateapj}
\usepackage{color}

\newcommand{\solar}{$_{\odot}$}
\newcommand{\tco}{$^{12}$CO}
\newcommand{\ttco}{$^{13}$CO}
\newcommand{\ceto}{C$^{18}$O}

\newcommand{\joz}{$J$=1$\rightarrow$0}

\newcommand{\kms}{\,km\,s$^{-1}$}
\newcommand{\degree}{$^{\circ}$}
\newcommand{\fdeg}{$^{\circ}$\hspace{-1mm}.}
\newcommand{\fmin}{$'$\hspace{-1mm}.}
\newcommand{\nf}{$\nu_N$}
\newcommand{\no}{$\nu_o$}
\newcommand{\tex}{$T_{\rm ex}$}
\newcommand{\tsys}{$T_{\rm sys}$}
\newcommand{\vlsr}{$V_{\rm LSR}$}
\newcommand{\tmb}{$T_{\rm mb}$}

\newcommand{\htwo}{H$_2$}

\def\lapp{\ifmmode\stackrel{<}{_{\sim}}\else$\stackrel{<}{_{\sim}}$\fi}
\def\gapp{\ifmmode\stackrel{>}{_{\sim}}\else$\stackrel{>}{_{\sim}}$\fi}

\shorttitle{ThrUMMS  I: Overview and First Results}
\shortauthors{Barnes et al.}

\begin{document}

\title{The Three-mm Ultimate Mopra Milky Way Survey. I. \\
    Survey Overview, Initial Data Releases, and First Results}


\author{
Peter J. Barnes\altaffilmark{1,2}, Erik Muller\altaffilmark{3}, Balthasar Indermuehle\altaffilmark{4}, Stefan N. O'Dougherty\altaffilmark{5}, \\
Vicki Lowe\altaffilmark{6,4}, Maria Cunningham\altaffilmark{6}, Audra K. Hernandez\altaffilmark{7}, and Gary A. Fuller\altaffilmark{8} \\
}
\email{pjb@astro.ufl.edu}


\altaffiltext{1}{Astronomy Department, University of Florida, P.O. Box 112055, Gainesville, FL 32611, USA}
\altaffiltext{2}{School of Science and Technology, University of New England, NSW 2351, Australia}
\altaffiltext{3}{National Astronomical Observatory of Japan, Chile Observatory, 2-21-1 Osawa, Mitaka, Tokyo 181-8588, Japan}
\altaffiltext{4}{CSIRO Astronomy and Space Science, P.O. Box 76, Epping, NSW 1710, Australia}
\altaffiltext{5}{College of Optical Sciences, University of Arizona, 1630 E. University Blvd., P.O. Box 210094, Tucson, AZ 85721, USA}
\altaffiltext{6}{School of Physics, University of New South Wales, NSW 2052, Australia}
\altaffiltext{7}{Astronomy Department, University of Wisconsin, 475 North Charter St., Madison, WI 53706, USA}
\altaffiltext{8}{Jodrell Bank Centre for Astrophysics, Alan Turing Building, School of Physics and Astronomy, University of Manchester, Manchester, M13 9PL, UK}

\begin{abstract}
We describe a new mm-wave molecular-line mapping survey of the southern Galactic Plane and its first data releases.  The Three-mm Ultimate Mopra Milky Way Survey (ThrUMMS) maps a 60\degree$\times$2\degree\ sector of our Galaxy's fourth quadrant, using a combination of fast mapping techniques with the Mopra radio telescope, simultaneously in the \joz\ lines of \tco, \ttco, \ceto, and CN near 112\,GHz at $\sim$arcminute and $\sim$0.3\kms\ resolution, with $\sim$2\,K\,channel$^{-1}$ sensitivity for \tco\ and $\sim$1\,K\,channel$^{-1}$ for the other transitions.  The calibrated data cubes from these observations are made available to the community after processing through our pipeline.  Here, we describe the motivation for ThrUMMS, the development of new observing techniques for Mopra, and how these techniques were optimised to the objectives of the survey.  We showcase some sample data products and describe the first science results on CO-isotopologue line ratios.  These vary dramatically across the Galactic Plane, indicating a very wide range of optical depth and excitation conditions, from warm and translucent to cold and opaque.  The population of cold clouds in particular have optical depths for \tco\ easily exceeding 100.  We derive a new, nonlinear conversion law from \tco\ integrated intensity to column density, which suggests that the molecular mass traced by CO in the Galactic disk may have been substantially underestimated.  This further suggests that some global relationships in disk galaxies, such as star formation laws, may need to be recalibrated.  The large ThrUMMS team is proceeding with several other science investigations. 
\end{abstract}

\keywords{astrochemistry --- galaxies: the Milky Way --- ISM: kinematics and dynamics --- ISM: molecules --- radio lines: ISM --- stars: formation}

\section{Introduction}\label{intro}
The Milky Way is a large, massive, and luminous disk galaxy, and since we are immersed in it, it can provide detailed and localised keys to understanding larger-scale processes that operate in other galaxies throughout the observable universe.  For example, our Galaxy uniquely enables parsec-scale opportunities to calibrate, and understand the origin and evolution of: the Schmidt-Kennicutt relations \citep{k98,k07,L08}, the radio-FIR correlation \citep{z10}, the conversion of HI into \htwo\ \citep{lg03,g07,v10}, the origin of turbulence in the interstellar medium \citep{b03}, and many other processes which are typically not resolved to better than $\sim$100\,pc scales in even nearby galaxies.

Therefore, high-resolution wide-field surveys of the Milky Way have an important impact on several of these topics.  For example, in studies of star formation, there is a critical threshold for physical resolution of about a parsec, since that is the relevant scale for the formation of individual clusters from dense molecular clumps, and these are likely the basic unit of star formation activity \citep{ll03} throughout the Galactic disk.  For a typical distance of 3\,kpc, the clump/cluster scale is only reached with angular resolutions of an arcminute or better.

At the largest scales, recent wide-field Galactic Plane surveys such as GLIMPSE \citep{bb03} and its successors, ATLASGAL \citep{s09}, and Hi-GAL \citep{m10} are revolutionising our understanding of the Milky Way, by enabling the {\em global} study of the cold molecular structures of the interstellar medium (ISM) and the star-forming activity within these structures, throughout the disk.  Most, if not all, of these surveys have been made feasible by the development of wide-format cameras or multi-pixel bolometers, enabling the coverage of hundreds of square degrees of sky at angular resolutions of only a few arcseconds.  While sensitively revealing the two-dimensional morphology and broadband spectral energy distributions (SEDs) of many objects, many of these surveys are in the continuum only; that is, they cannot provide kinematic nor line-of-sight distance information.

\notetoeditor{}
\begin{figure*}[t]
\centerline{\includegraphics[angle=0,width=179mm]{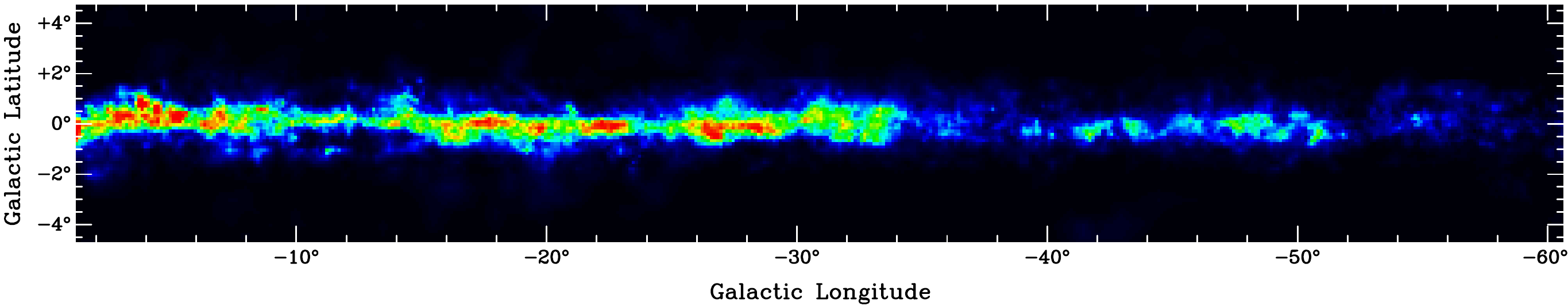}}
\vspace{-21mm}
\begin{picture}(440,10)
{\color{red}
   \thicklines
   \put(33.4,2.1){\framebox(471,16){}}
} 
\end{picture}
\vspace{18mm}
\caption{Columbia-CfA \tco\ survey $W_{CO}$ image of the Milky Way's 4th Quadrant between longitudes 359\degree\ and 299\degree\ \citep{dht01}.  The black-to-red colour scale is over 0--200\,K\kms.  The red box shows the original planned coverage of ThrUMMS.  As of 2014 Dec, mapping of this region is $\sim$80\% complete, yielding $\sim$7$\times$ higher resolution maps than the CfA survey ($\sim$50$\times$ higher areal resolution), and simultaneously in the 4 molecular lines \tco, \ttco, \ceto, and CN.
}
\label{cfa}
\end{figure*}

Comprehensive spectral-line surveys with this level of wide-field coverage are rarer and much more limited: the main wide-field surveys of note are in HI \cite[e.g., the SGPS in the south;][]{m05}, \tco\ \cite[e.g., the Columbia-CfA survey;][]{dht01} or \ttco\ \cite[the GRS in the north;][]{j06}, and more recently \htwo\ \cite[UWISH2 in the south;][]{f11}.  Fewer still have adequate spatial resolution across these wide scales.  So while detailed surveys of portions of the Galactic ISM exist, each of these necessarily has its set of limitations.

Thus, UWISH2 lacks a practical velocity resolution for comparison to the radio-frequency surveys, and focuses on only one spectral line, limiting its power as a physical diagnostic.  The SGPS is wide-field, but the velocity and angular resolutions are too modest (1\kms, 2$'$) for detailed work involving correlation and comparison of the numerous and significantly higher resolution near- to far-IR datasets mentioned above.  Further, the HI mapped by such surveys comprises a different phase of the ISM than that directly involved in star formation.  The venerable Columbia-CfA survey (Fig.\,\ref{cfa}) is at a resolution of only 8$'$, or $\sim$7\,pc for a typical ``nearby'' (i.e., distance $<$ 3\,kpc) molecular cloud, and is only in one species.  The Nanten \tco\ survey is only slightly higher resolution (4$'$), still well above the clump resolution threshold.  
The GRS \citep{j06} is at a reasonable resolution (0\fmin8), but only covers 37\degree$\times$1\degree\ of the first quadrant, and again is in just one species (\ttco).  CHaMP \citep{b11} covers 120\,deg$^2$ of the fourth quadrant and is multi-species and reasonable resolution (0\fmin6), but the fields mapped are of moderate width ($\sim$1\degree\ or less) and focused on the denser molecular clumps.  There are also a few other $\sim$degree-wide molecular cloud maps in the fourth quadrant recently made at Mopra, but their areal coverage is substantially less than CHaMP's.

Thus, by various metrics, we have no complete, wide-field, multi-species inventory of the gaseous content of the ISM at an angular resolution approaching that (20$''$ or better) of the continuum surveys.  Future large-scale Galactic ISM surveys, such as GASKAP \citep{d13}, will soon yield HI maps with resolutions of 20$''$ and 0.1\kms\ and high sensitivity, providing a significant improvement over the existing SGPS and its sister HI surveys.  But to explore the molecular content of the Galaxy, we must still contend with some fundamental observational limitations.

ThrUMMS' genesis occurred in 2009 at a meeting of the Mopra community (the Millimetre Astronomy Legacy Team, or MALT), where a need for three large legacy projects was identified, each covering the southern Galaxy in a different frequency range and having related, but somewhat different, science goals (see below).  The 110\,GHz project (subsequently dubbed ThrUMMS by its participants, and also called MALT-110) was designed to address the wide-field issues described above.

In this paper we introduce the {\em Three-mm Ultimate Mopra Milky Way Survey} (ThrUMMS), and provide the background information on ThrUMMS for the community to make best use of ThrUMMS' wide-field, molecular transition data products.  In the next section we describe the survey design and the unique operational aspects of the data collection and reduction.  In \S\ref{results} we describe the current Data Releases and the data products they contain.  \S\ref{disc} presents our first science results for the global dataset. 
\S\ref{concl} concludes with a brief summary.

\section{Survey Design}\label{design}

\subsection{The Need for Faster Mapping Techniques}\label{VFM}
Even today, most radio telescopes have single-pixel detectors in their focal plane.  For single-dish antennas operating over the $\sim$1--100\,GHz range, angular resolutions are typically limited to about an arcminute.  Without heterodyne focal-plane detector arrays (which are still rare and have most recently tended to focus on narrow-field astrochemical studies) to speed up the mapping, even this low resolution takes an extraordinary time to completely map large areas of sky at reasonable sensitivity.  The conventional wisdom for some time has been that the dedicated, small-telescope, CfA/Nanten model was the only practical way to produce useful wide-field molecular maps.

Motivated by this conundrum, we devised two new techniques specifically optimised for the 22\,m diameter Mopra\footnote{Operation of the Mopra radio telescope is made possible by funding from the National Astronomical Observatory of Japan, the University of New South Wales, the University of Adelaide, and the Commonwealth of Australia through CSIRO.} 
antenna, to dramatically improve the telescope's speed for wide-field mapping.  We also implemented a third technique (pioneered in the CfA survey) that improves the image fidelity for all mapping projects.  A detailed exposition of these techniques has been given by \citet{bd09}, but for completeness we highlight the essential features here (summarised in Table \ref{features}).

\notetoeditor{}
\begin{figure*}[t]
\centerline{
\includegraphics[angle=0,scale=0.65]{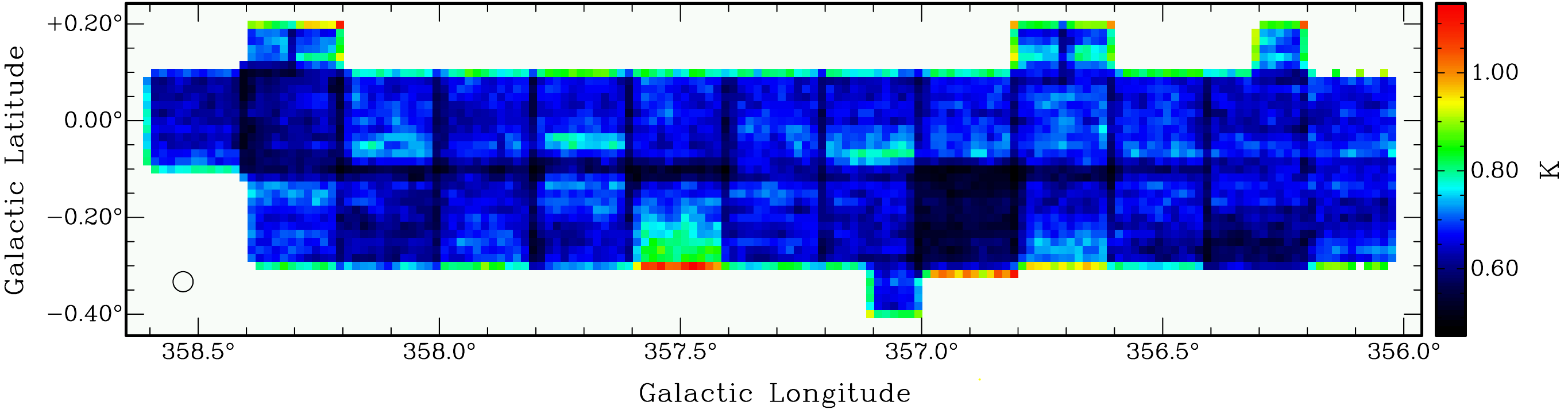}
}
\vspace*{4mm}
\hspace*{94mm}
\includegraphics[angle=0,scale=0.45]{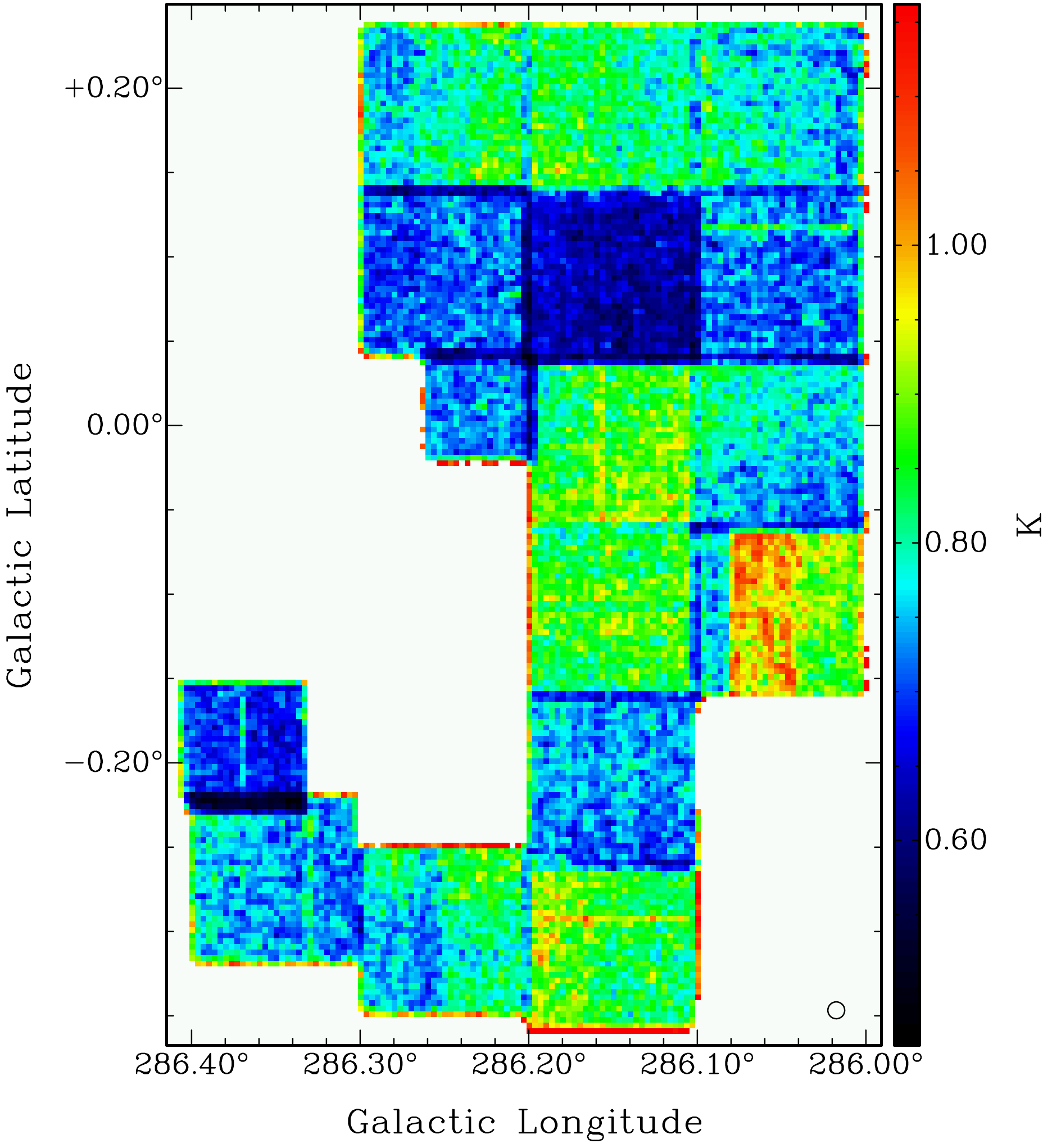}

\vspace*{-95mm}\hspace{-1mm}
\parbox{86mm}{
\caption{RMS noise maps for a combination of BSM and AM (above, from Dame et al., in prep.) and normal OTF mapping with a fixed \nf\ = \no\ \citep[right, from][]{b11}, each over a large number of OTF fields (26 \& 16 resp.).  The effective beam sizes for the different mapping techniques are shown in the lower-left and lower-right corners, resp.  Both maps are shown on the same brightness scale to make the comparison more straightforward.  For the AM/BSM map, there is only one modal peak in the rms at 0.65\,K, due to the sky-active setting of \nf.  For the normal OTF map, the modal rms is 0.75\,K, but there is a much larger variation in rms than in the AM/BSM map due to the rms' strong elevation dependence.  There are also blended secondary peaks in the histogram (shown in Fig.\,\ref{rmshisto}) at 0.85\,K due to poor weather for one OTF field, and at 0.65\,K due to excess coverage for another field.
\label{rmsmaps}}} 
\vspace*{3mm}
\end{figure*}

\subsubsection{Fast Mapping with Mopra\label{FM}}
The first of these techniques, ``Fast Mapping'' (FM), takes advantage of the fact that, especially for the \tco\ line but also for \ttco, the S/N performance enabled by a 22\,m diameter dish is very high, and for survey mapping of such bright lines, sensitivity requirements can be significantly relaxed compared to, for example, those needed for the common ``dense gas tracers'' such as HCN or CS.  For the latter, at Mopra the sampling interval of 2.0\,s coupled with Nyquist-sampled on-the-fly (OTF) mapping produces maps with typical rms noise of 0.3\,K\,ch$^{-1}$, which is well-suited to mapping of such lines with characteristic brightness $\sim$1--3\,K.  However, in the radiometer equation, the rms noise $\sigma_{\rm rms} \propto t^{-1/2}$ for integration time $t$.  So for example, a factor of 3 worse sensitivity, i.e., $\sigma\sim$ 1\,K\,ch$^{-1}$, for a very bright line (5--20\,K or more) dramatically reduces the integration time necessary to achieve a still excellent S/N, by (in this example) a factor of 3$^2 \sim$ 10.  In other words, with suitable modifications to the observing hardware and software, we can significantly reduce the sampling interval in a given observing mode, such as OTF mapping, and allow maps of bright lines to be completed perhaps an order of magnitude faster than with regular ``slow'' mapping.  In practice, the FM sampling interval was fixed to be about 8$\times$ faster than the regular sampling interval, namely 0.256\,s instead of 2.000\,s.

\subsubsection{Beam Sampled Mapping with Mopra\label{BSM}}
The second technique also takes advantage of a particular feature of Galactic \tco\ emission: it is spatially very widespread.  Again, for a survey mapping project, Nyquist sampling of the \tco\ emission (especially at the more modest sensitivity of FM) can be considered slightly superfluous.  Conveniently, the OTF mapping software at Mopra has an easily-set parameter which controls the degree to which the spatial sampling matches (or doesn't match) the Nyquist-sampling criterion, called the ``Nyquist frequency,'' \nf.  Thus, when \nf\ = \no\ (the 

\hspace*{10mm}.

\vspace*{42mm}\hspace*{-3mm}observing frequency), sampling during regular OTF mapping is set by default to be $\sim$10\% finer than the formal Nyquist spacing at \no.  For ThrUMMS, however, we set \nf=\no/2, so that OTF mapping in this case is performed with the telescope scanning the sky twice as fast as the default.  This effectively means that ThrUMMS maps are sampled at $\sim$half the Nyquist rate (i.e., in a spatially coarser way), in both the scanning direction and between scan rows.  With this change, OTF mapping obtains effectively beam-spaced (33$''$ at 115\,GHz) samples, a method we have dubbed ``Beam Sampled Mapping'' (BSM).  This follows a long tradition of ``complete'' but beam-spaced mapping from the early days of radio astronomy, including parts of the CfA survey itself \citep{dht01}.

Because this is a rather novel mapping approach, we describe here some aspects to BSM that, while not meaningfully affecting the data quality, are worth bearing in mind for some applications, e.g., those that depend on accurate structural information at the smallest effective resolution of the data.

\begin{figure*}[ht]
\centerline{
\includegraphics[angle=0,scale=0.4]{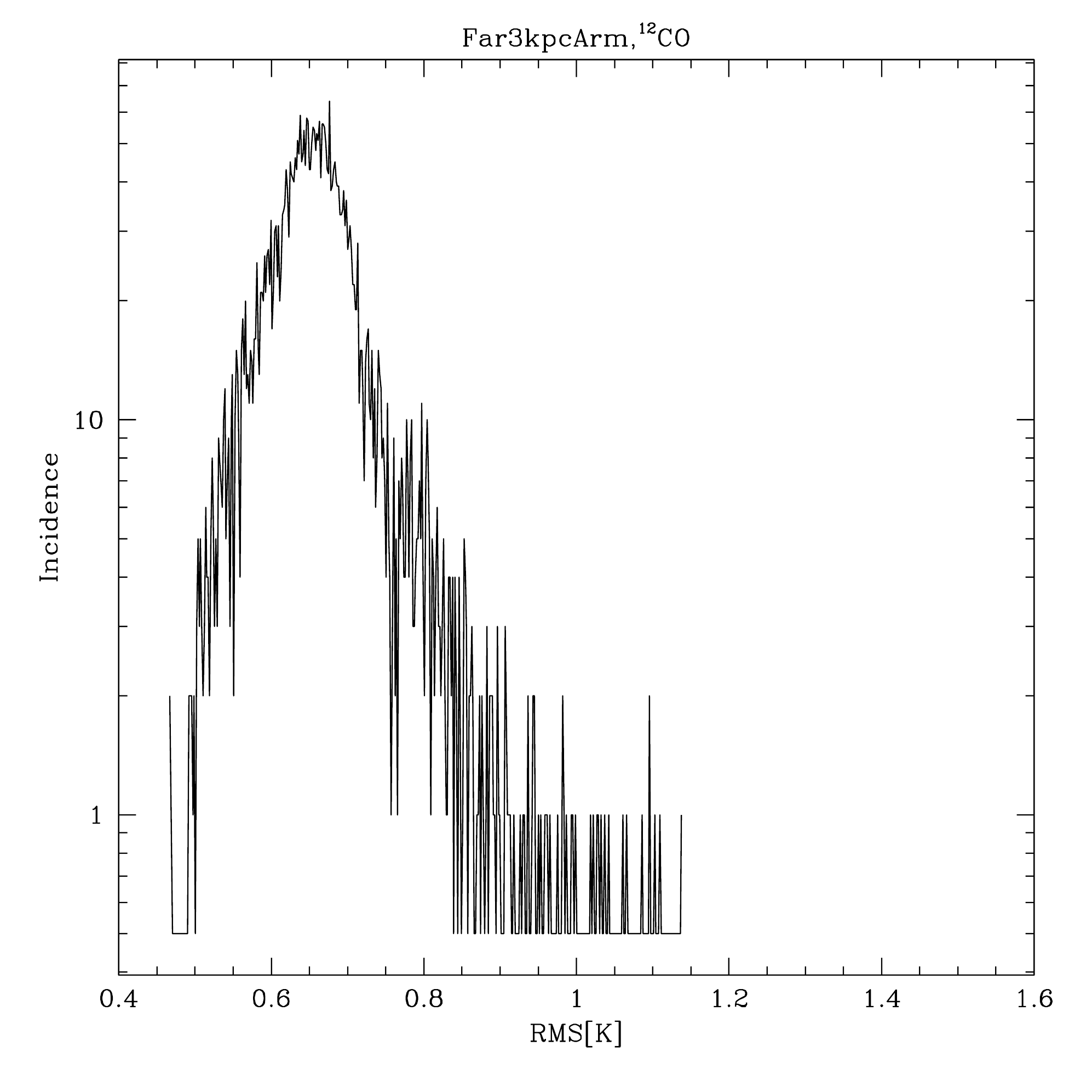}
\includegraphics[angle=0,scale=0.4]{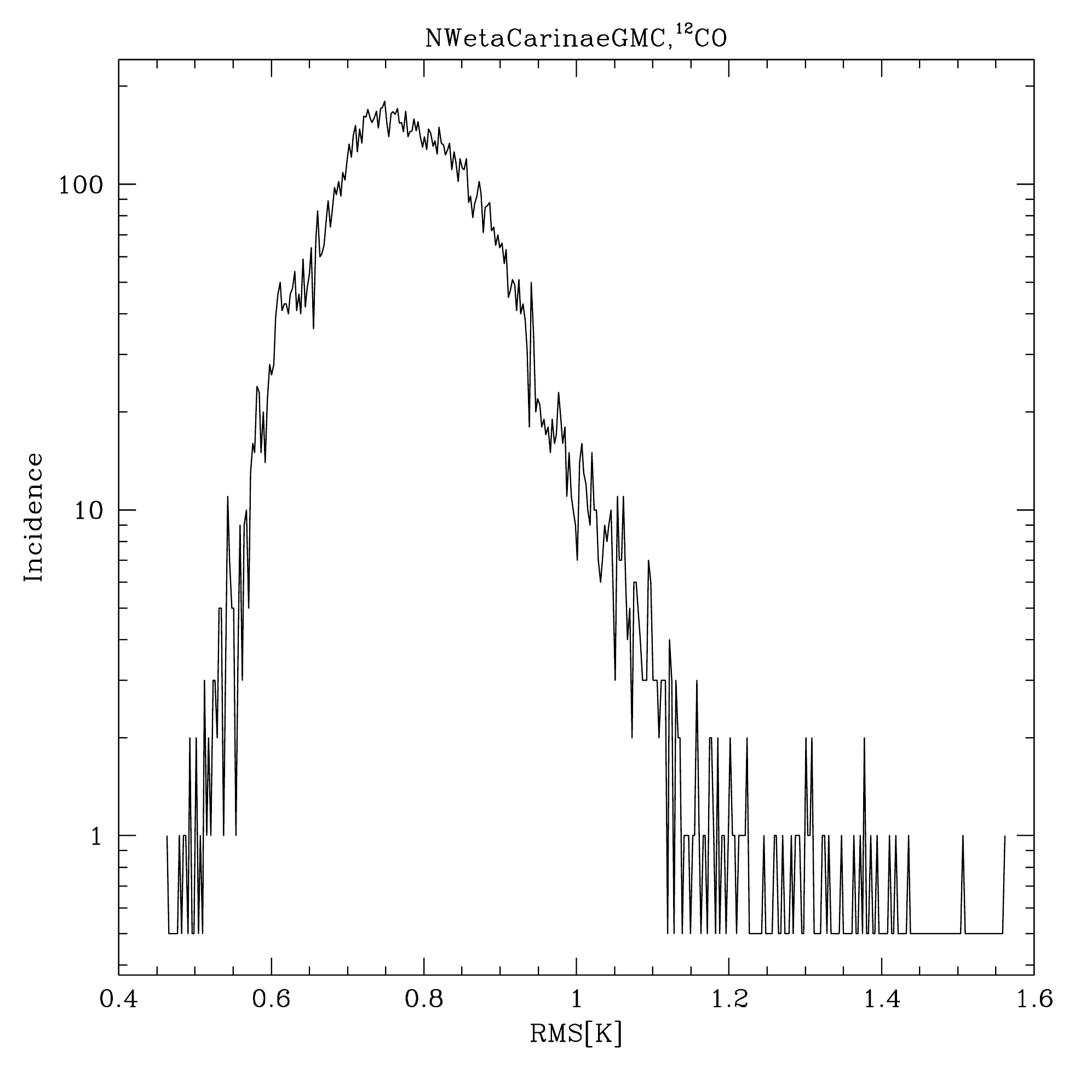}
}
\vspace{-3mm}
\caption{RMS noise histograms (log scale for $y$, same linear brightness scale for $x$) for the AM/BSM (left) and normal OTF (right) maps shown in Fig.\,\ref{rmsmaps}.
The narrower spread of system noise in the left panel is due to the noise-homogenisation effects of the Active Mapping technique.  The slight difference in normalisation of these plots is explained in the caption to Fig.\,\ref{rmsmaps}.
\label{rmshisto}}
\vspace*{-3mm}
\end{figure*}

\begin{deluxetable*}{llccccc}
\tabletypesize{}
\tablecaption{Mopra mapping techniques\label{features}}
\tablewidth{0pt}
\tablehead{
\colhead{Mapping} & \colhead{Sampling} & \colhead{Finest} & \colhead{Atmospheric Stability} & \colhead{Relative} & \colhead{Relative} & \\
\colhead{Mode} & \colhead{Interval} & \colhead{Resolution} & \colhead{Accommodated} & \colhead{Mapping Speed\tablenotemark{a}} & \colhead{RMS Noise\tablenotemark{a}} & \vspace{-3mm} \\
}
\startdata
regular & 2.000\,s & 33$''$ & cannot adjust & 1 & 1 (var.) &  \\
FM & 0.256\,s & 33$''$ & cannot adjust & 8 & $\sqrt{8}$ (var.) &  \\
BSM & any & 66$''$ & cannot adjust & 4 & 1 (var.) &  \\
AM & any & any & $>$1\,hr & $\leq$1 & 1 & \vspace*{1.8mm}  \\
ThrUMMS & 0.256\,s & 66$''$ & $>$1\,hr & $\leq$32 & $\sqrt{8}$ & \vspace*{-3mm}  \\
\enddata
\tablenotetext{a}{These factors are multiplicative: for ThrUMMS, all of FM+BSM+AM are used. 
\vspace*{2mm}}
\end{deluxetable*}

During normal OTF mapping, the antenna is continuously slewed across a raster line in the eventual OTF map.  During the slew, data are accumulated almost continuously in the MOPS digital filterbank, and interrupted very briefly each sampling interval (normally 2.0\,s) to dump out the contents of the spectrometer to mass storage.  The speed of the slew is tuned to the sampling inter- val so that, for normal OTF mapping, Nyquist sampling on the sky (at the given observing frequency) is achieved by default (i.e., \nf=\no).

During BSM+FM, even though the sampling interval is now about 8 times faster than for regular mapping, the telescope is typically slewing across the map 16 times faster, because \nf=$\frac{1}{2}$\no.  Thus, the data dumps from the spectrometer occur about once per beam in the raster direction.  Because the integration is continuous over this time, we are effectively forming a properly-smoothed average spectrum along the raster direction at $\sim$beam-spaced positions, and apart from this spatial convolution, there is no loss of information (such as zero-spacing data) on source structure on these scales.

Across the raster lines, however, the situation is different.  Here the selection of \nf=$\frac{1}{2}$\no\  means that the raster lines are separated by one intrinsic Mopra beamsize, and we are therefore undersampling the sky emission in this direction.  In other words, we could be missing some zero-spacing information in this dimension.  However, Mopra's intrinsic angular resolution of 0\fmin6 at 3\,mm is well below the resolution threshold for cluster-formation (see \S\ref{intro}), so even with beam-spaced sampling, this issue is not of particular concern for ThrUMMS science.  For example, in some planned science investigations, ThrUMMS maps are intended to be starting points for more detailed follow-up studies.  More importantly, we allow for this feature of the mapping in our data pipeline, producing reliable maps of the emission at the quoted effective beam size (see \S\ref{reduction}).

With BSM, since we do not change any of the parameters in the radiometer equation, we pay no noise penalty {\em per se} for such maps.  Obviously, however, the maps so generated are at twice Mopra's inherent angular resolution.  But mapping a given area now proceeds four times faster than before, which for ThrUMMS we consider a very valuable trade-off.  In summary, the net effect of combining the FM and BSM techniques results in maps of sensitivity $\sim$1\,K\,ch$^{-1}$ at an angular resolution of $\sim$1\fmin2, but made at an areal mapping speed roughly 32$\times$ faster than Nyquist-sampled ``slow'' OTF mapping.  This is a dramatic improvement over (e.g.) the CfA or Nanten maps, and ideal for our purposes.  We call this combination of techniques ``Very Fast Mapping,'' or VFM.

\begin{deluxetable*}{cclccccl}
\tabletypesize{}
\tablecaption{ThrUMMS 3mm band spectral lines\label{linepars}}
\tablewidth{0pt}
\tablehead{
\colhead{Species} & \colhead{Transition} & \colhead{Frequency} & \multicolumn{2}{c}{Efficiencies\tablenotemark{a}} & \colhead{Resolution\tablenotemark{b}} & \colhead{RMS\tablenotemark{c}} & \colhead{Science Topics} \\
 &  & \colhead{(GHz)} & \colhead{$\eta_b$} & \colhead{$\eta_c$} & \colhead{(\kms)} & \colhead{(K)} & \vspace{-3mm} \\
}
\startdata
\tco & \joz & 115.271 & 0.42 & 0.55 &0.33&1.3&largest scales, GMC envelopes, temperatures, outflows \\
CN & \joz & 113.5 (5 hf)\tablenotemark{d} & 0.42 & 0.56&0.33&0.9&dense clumps, Zeeman candidates \\
\ttco & \joz & 110.201 & 0.43 & 0.57 & 0.34 & 0.7&PDRs, fractionation, optical depth, velocity dispersion \\
\ceto & \joz & 109.782 & 0.43 & 0.57 & 0.34 & 0.7 & column density, depletion, dense gas finder \vspace*{-3mm} \\
\enddata
\tablenotetext{a}{Following \citet{ku81}, the beam efficiency $\eta_b$ is appropriate for obtaining \tmb\ of compact ($\sim$beam-sized) sources, while the coupling efficiency $\eta_c$ is appropriate for computing \tmb\ of more extended (several arcmin) sources.}
\tablenotetext{b}{This is for a 4-channel binning: the intrinsic velocity resolution is 0.085\kms.}
\tablenotetext{c}{This is for a 4-channel binning: the intrinsic rms noise per full-resolution channel is twice as large.}
\tablenotetext{d}{This line has 9 hyperfine components, of which 5 are recorded in the IF zoom band.\vspace*{1mm}}
\end{deluxetable*}

\subsubsection{Active Mapping with Mopra\label{AM}}
The third technique, new to Mopra but used at some other facilities, is a sky-related modification of BSM, but which can also be applied to regular ``slow'' OTF mapping.  For most modern mm-wave radio telescope systems, the system temperature $T_{\rm sys}$ is dominated by the contribution from the atmosphere, $T_{\rm atm}$.  Apart from weather-related factors such as humidity, this of course depends on the airmass at the elevation $e$ of the observation above the horizon, $T_{\rm atm}\propto$ cosec($e$).  It turns out that we can easily (and exactly) accommodate the predictable, elevation-dependent component of $T_{\rm sys}$ by scaling \nf\ to $T_{\rm sys}$($e$)/$T_{\rm sys}$(90\degree), since the slightly slower mapping at a larger \nf\ will increase the integration time in exact proportion to that required to adjust for the higher $T_{\rm sys}$($e$) compared to $T_{\rm sys}$(90\degree).  Thus,
\begin{eqnarray}
	\nu_{N}(e) & = & \nu_{N}(90^{\circ}) \left(\frac{T_{\rm sys}(e)}{T_{\rm sys}(90^{\circ})}\right)   ,
\end{eqnarray}
with the minimum \nf(90\degree) chosen to be 55\,GHz.  We call this technique ``sky-active mapping'', or AM.

In practice, the minimum $T_{\rm sys}$(90\degree) for the scaling of \nf\ was chosen to be a conservative approximate global minimum in the best conditions (400\,K at 115\,GHz in our case), rather than estimated separately for each day at $e$=90\degree.  In that case, slowly-varying ($\sim$hourlong) and day-to-day weather changes could also be accommodated by this scheme just as easily.  Only in the most variable weather is the AM method susceptible to bad spectral baselines, producing ``weather stripes'' in the maps, but this is true for any OTF mapping project, and the weather in such cases is, on average, not good enough to observe at all.  This sky-active technique at least evens out much of the weather variation, and usually the scheme works quite well: an illustrative example of the combination of two of these techniques (AM+BSM) using Mopra's OTF system is shown in Figures 2 
and \ref{rmshisto}, and is compared with a similar mosaic made with regular OTF mapping.

\subsection{Spectroscopy\label{lines}}
Mopra gives us another important advantage over mapping surveys made elsewhere: the MOPS spectrometer.  This further combination makes ThrUMMS rather unique compared to previous molecular surveys of the Milky Way.

\begin{figure}[b]
\includegraphics[angle=0,scale=0.34]{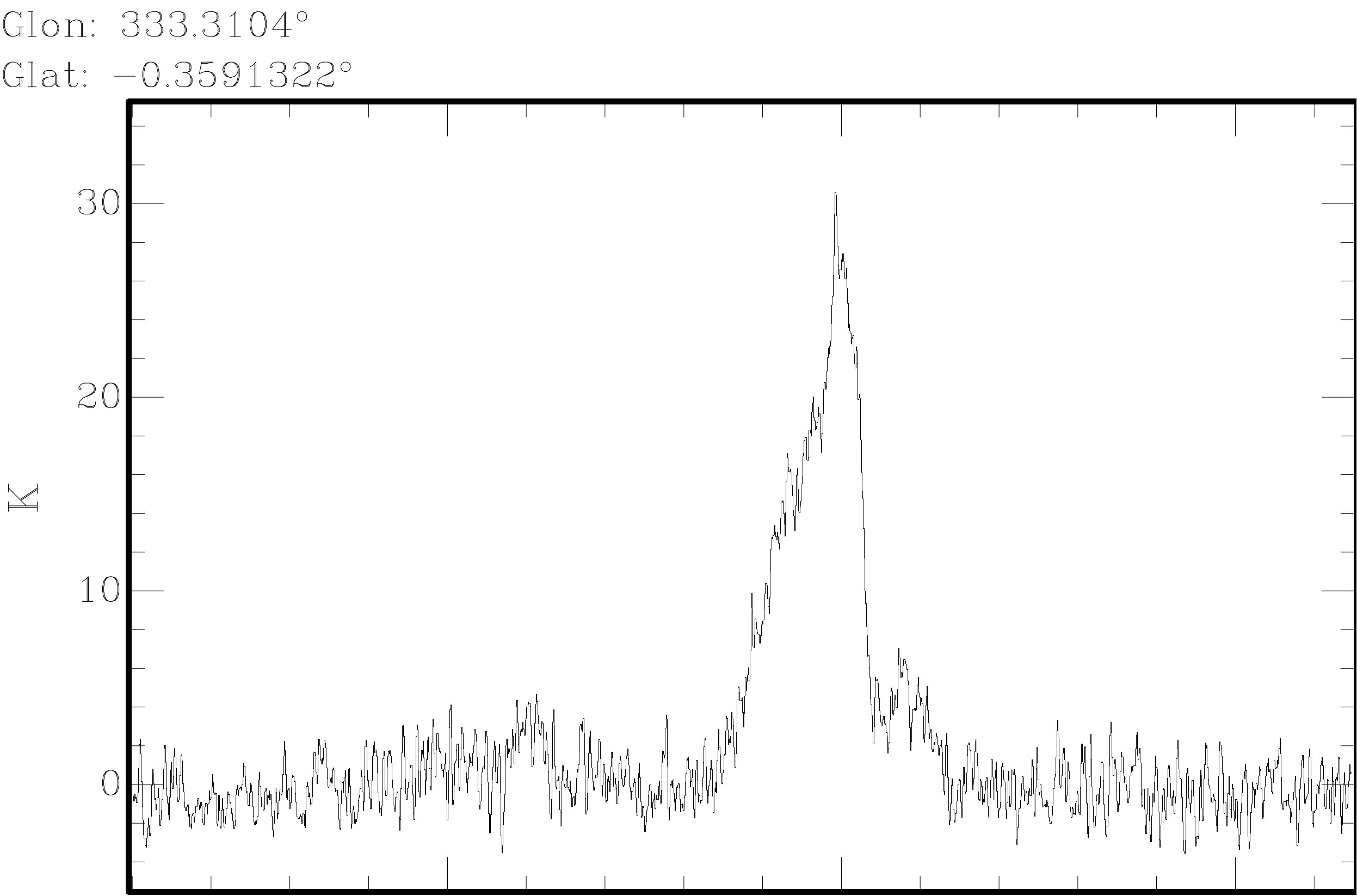}

\vspace*{-0.5mm}\includegraphics[angle=0,scale=0.34]{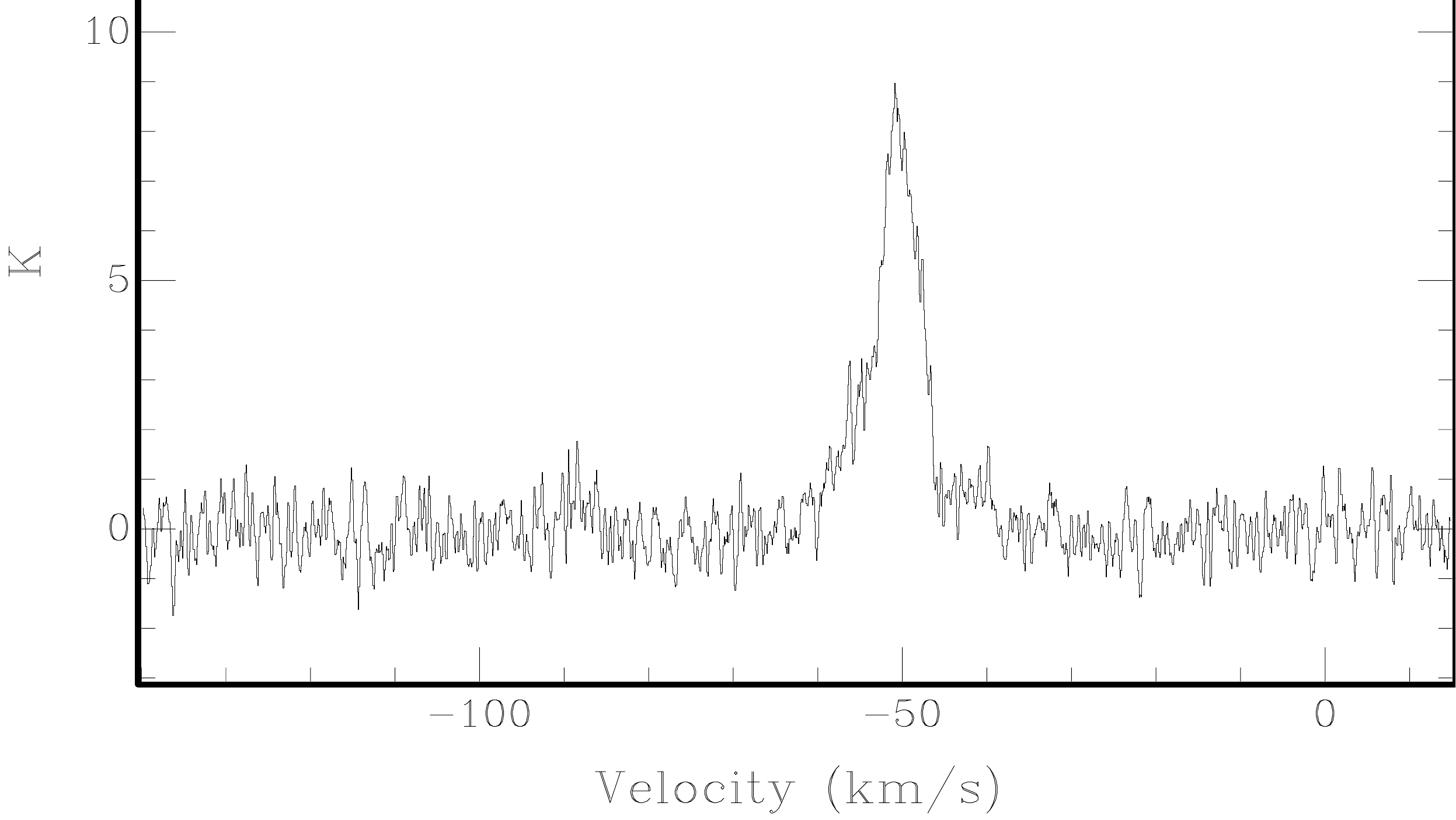}

\vspace*{-92mm}\hspace*{15mm}\parbox{26mm}{\Large \tco}

\vspace*{45mm}\hspace*{15mm}\parbox{26mm}{\Large \ttco}
\vspace*{36.5mm}
\caption{Sample spectra from ThrUMMS data cubes.  After a 4-channel binning (changing the intrinsic 0.085\kms\ channel width to 0.34\kms), the per channel rms noise in each spectrum is $\sim$1.3\,K (\tco, top) and $\sim$0.7\,K (\ttco, bottom).  Note also that we typically have good baselines by choosing OFF positions at similar elevations to the observed maps.
}
\label{spectra}
\end{figure}

For ``slow'' OTF mapping, MOPS simultaneously records up to 16 high spectral resolution ($\sim$0.1\kms) IF ``zoom modes'' of width 137\,MHz each ($\sim$400\kms\ at 3\,mm), which together can semi-arbitrarily cover an 8\,GHz wide portion of the 3\,mm band.  Especially in the line-rich 90\,GHz range, Mopra can therefore map 16 or more spectral lines extremely efficiently, as in CHaMP \citep{b11} and other surveys of dense molecular cores and clumps.  During FM, however, the data transfer rate from the spectrometer hardware to mass storage is maximised, and imposes a limit of 4 IF zooms on the data-taking.  This limitation, however, does not affect the science goals of ThrUMMS, since there are only 4 transitions bright enough (at the sensitivity of FM) with a practical astrophysical application in the 107--115\,GHz range.  These transitions are listed in Table \ref{linepars}.  We show in Figure \ref{spectra} some sample spectra 
to illustrate the spectral data quality during VFM+AM.

\subsection{Coverage of the Galactic Plane}

In 2012 we completed (except as noted below) all mapping for the equatorial degree-wide strip (i.e., $|b| <$ 0\fdeg5) for two-thirds (360\degree\ $>l>$ 300\degree) of the Fourth Quadrant (hereafter 4Q).  In 2013 we began expanding the coverage to $|b|<$ 1\degree\ over the same longitude range, and anticipate that this will be completed in 2015.  
While optimising the data-taking for efficiency and low noise, the filling of our 60\degree$\times$2\degree\ target area during the observing is somewhat random, but large contiguous portions of the 4Q are now available.  A small number of fields are still affected by poor weather or equipment problems, such that they will require re-observation to maintain uniform sensitivity; other fields were not finished in one session.  Despite this patchwork progress, there are no calibration ``edges'' between fields observed at different times or seasons, since the calibration is robust (see \S\ref{obsprocs} and \S\ref{reduction}).  
Moreover, because of our variable-\nf\ approach to noise management, the rms varies only modestly between fields, with very little (if any) effect of such blended data being apparent in the merged data cubes.

\begin{figure}[t]
\hspace{0mm}\includegraphics[angle=0,scale=0.54]{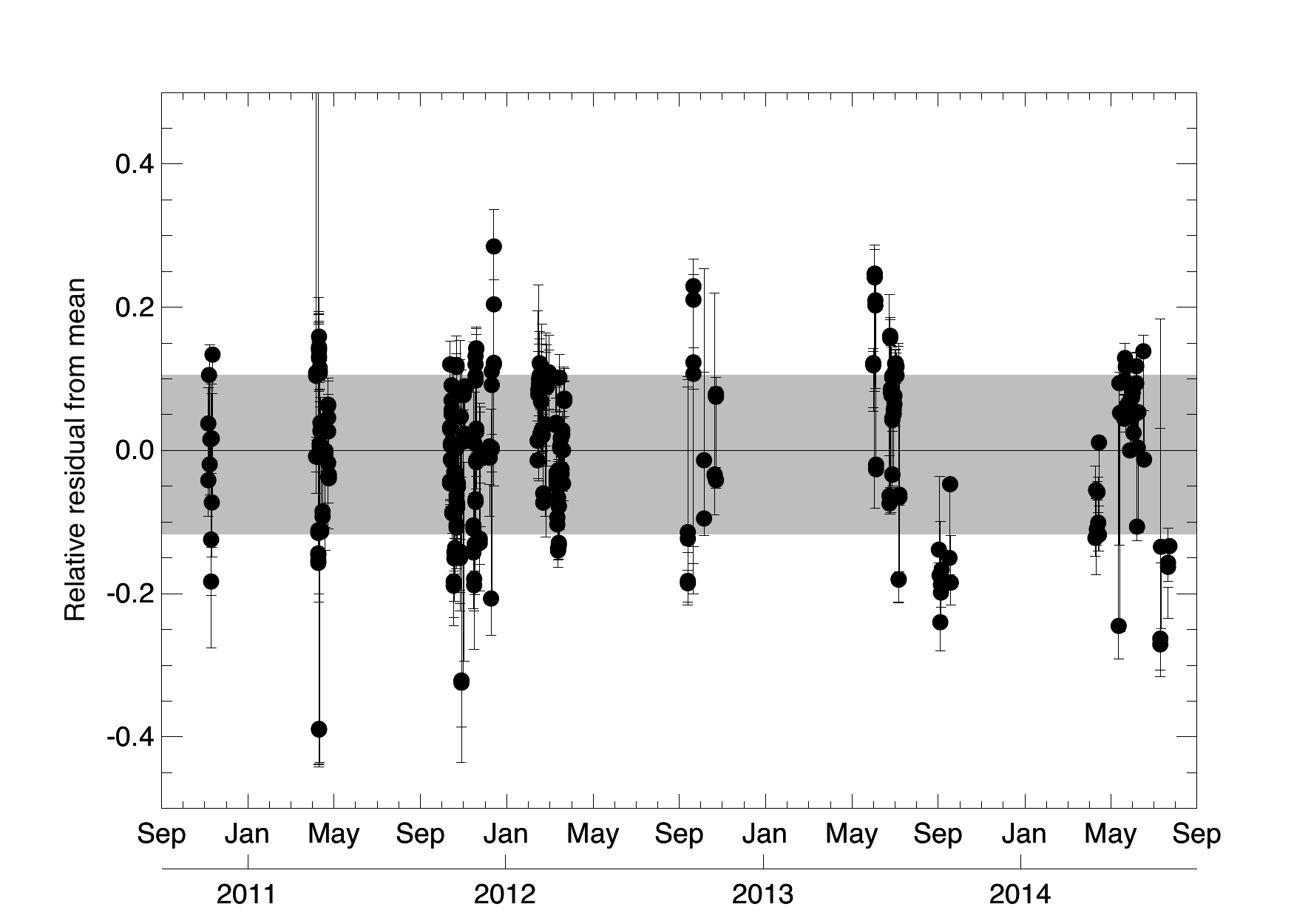}
\caption{Calibration data from all five ThrUMMS observing ``seasons'' to date, on the source BYF99a \citep{b11}.  Each point shows a fit to the emission in this source for the calibration scan at the beginning of each day's observing, on an ordinate scale normalised to the mean of all measurements.  Shown in grey is the mean $\pm$ standard deviation of all these fits; the normalised standard deviation is 11\%.
}
\label{cals}
\end{figure}

\subsection{Observing Procedures\label{obsprocs}}

All ThrUMMS maps are made with the VFM OTF raster direction in longitude (i.e., raster lines are at constant latitude).  The absolute pointing accuracy is monitored and corrected to within 6$''$ using various 86\,GHz SiO maser sources \cite[R Car, IRSV\,1540, AH Sco, VX Sgr;][]{i13} between each map.  Pointing variations rarely exceed 20$''$ at the end of each OTF map.  (With the 24$''$ pixel size in each data cube, such pointing corrections, although slightly larger than those for smaller OTF maps at Mopra, are inconsequential here.)  At the beginning of each daily observing session, a calibration scan is taken of the bright molecular clump BYF\,99 in the $\eta$ Carinae Nebula complex (Fig.\,\ref{cals}), and the $T_{\rm sys}$ measurements made to start the \nf\ scaling for subsequent VFM.  OFF positions for each VFM field are taken from the list of \citet{dht01}.  Under good conditions each 0\fdeg5 field takes about 2$^h$ of clock time, or somewhat longer as the $T_{\rm sys}$ rises.  Fields are prioritised for mapping so that, whenever possible, they start at an hour angle (HA) near --1$^h$, and finish near HA = +1$^h$, minimising the airmass and maximising the mapping efficiency.  This procedure is controlled by a robotic telescope observing script, implemented in the control software since 2010, and requiring only occasional oversight functions from our observing team.  

\notetoeditor{}
\begin{figure}[b]
\includegraphics[angle=0,scale=0.45]{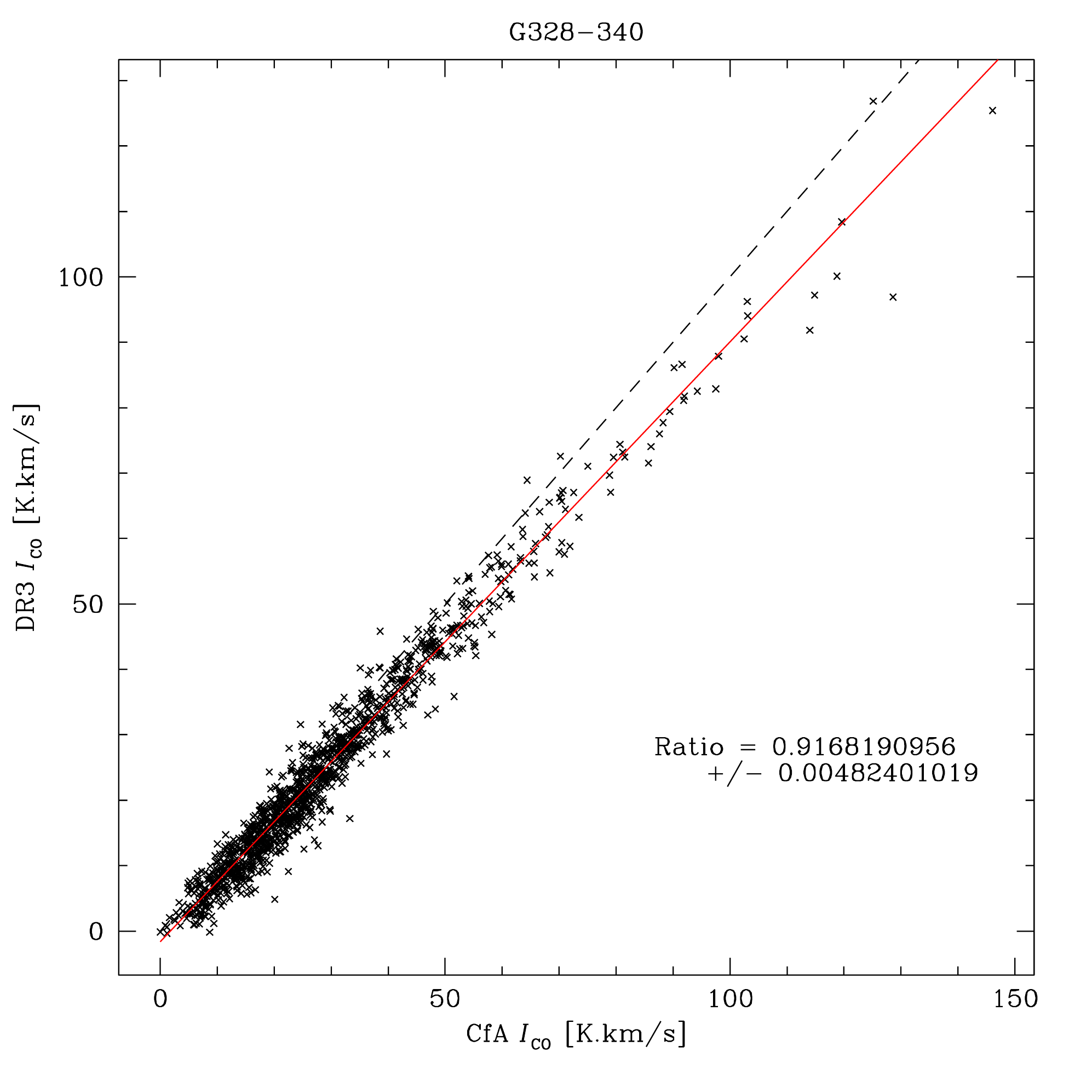}

\vspace{-1.0mm}
\caption{Integrated intensities from DR3 \tco\ maps covering 340\degree$>$$l$$>$328\degree and integrated over --55$<$\vlsr$<$--40\,\kms, compared with the same measurements from an equivalent area in the CfA survey \citep{dht01}.  The DR3 data were convolved to the CfA beam of 8\fmin4 to form this pixel-by-pixel comparison.  (This is the same area and velocity range as displayed in the 2nd and 3rd panels of Fig.\,\ref{comps}.)  Points from pixels at the edge of the ThrUMMS comparison area, where the convolution to the larger CfA beam becomes inaccurate, are excluded from this plot.  The dashed line indicates a 1:1 equivalence; the average result over this and several other areas and velocity ranges is given by the red line at 90$\pm$5\%.  This indicates that the ThrUMMS data are on a brightness scale slightly below that of the CfA data, but that the relative consistency between the two surveys is excellent, vindicating our quality-control procedures.
}
\label{cfacal}
\end{figure}

\subsection{Data Reduction and Processing\label{reduction}}

We use an augmented version of the \textsc{Livedata/Grid-zilla} package \citep{b01}, the usual standard for ATNF single-dish data reduction.  This package provides a convenient environment for basic spectral calibration, baselining, separation of different IFs and spectral line data, gridding and convolution, and the formation of standard FITS ($l,b,V$) cubes.  We describe our procedures next, including custom improvements to the package.

\notetoeditor{}
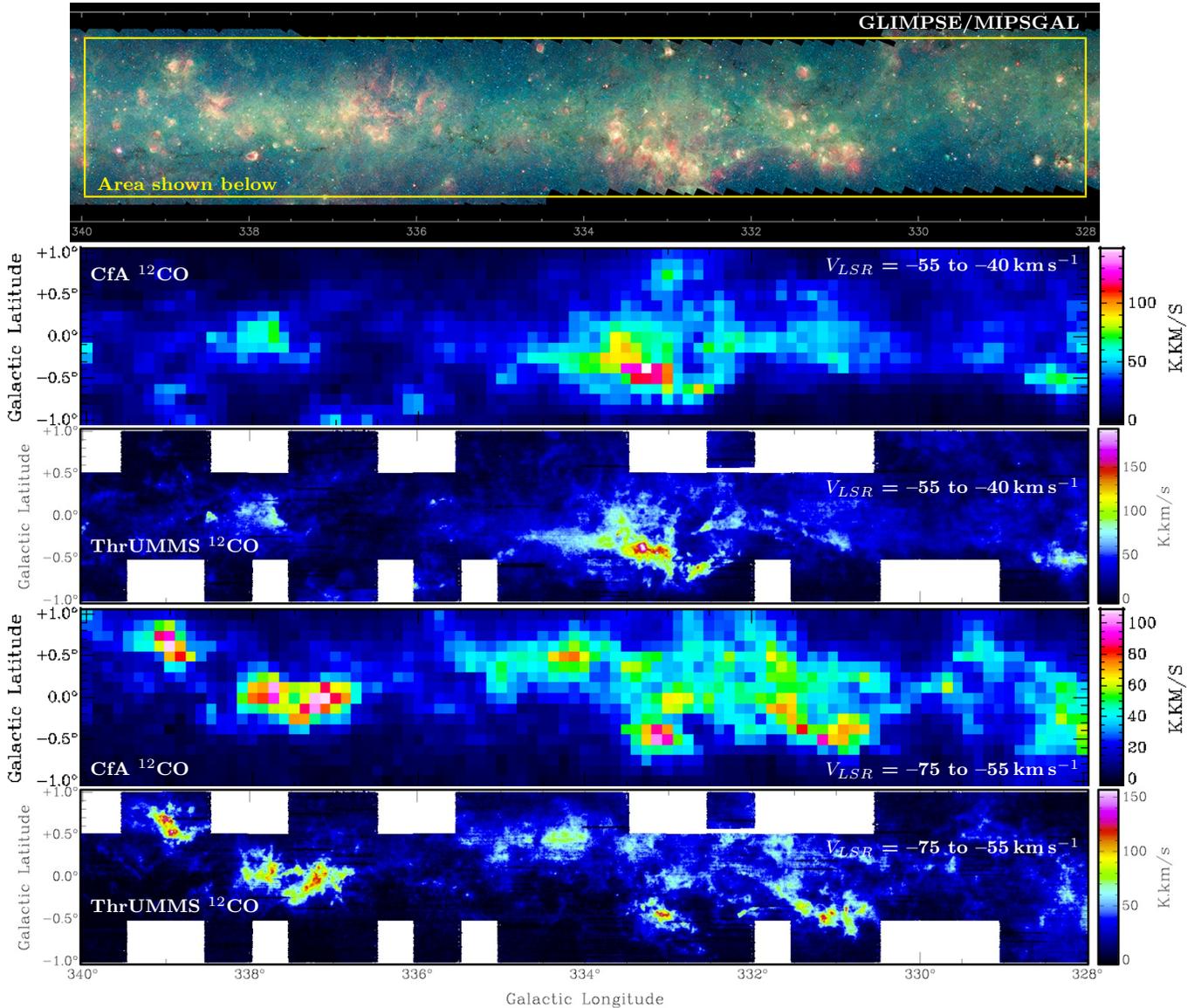
\begin{figure*}[t]
\hspace{10mm}\includegraphics[angle=0,scale=0.337]{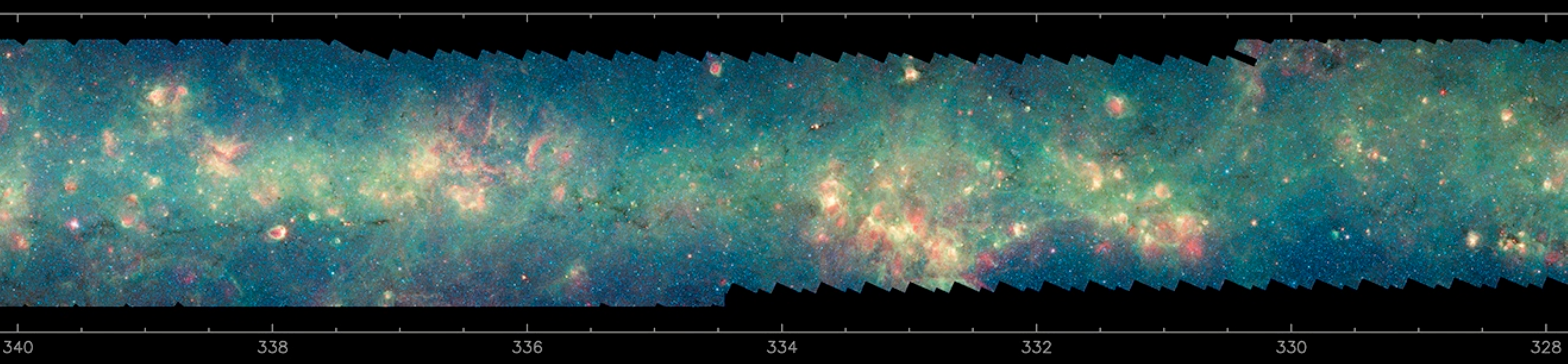}
{\color{yellow}
   \begin{picture}(447,10)
   \thicklines
   \put(35,31){\framebox(433,68){}}
   \end{picture}
   
   \vspace{-15mm}\hspace{14mm}\parbox{46mm}{\bf Area shown below}
} 

\vspace{8mm}\includegraphics[angle=0,scale=0.65]{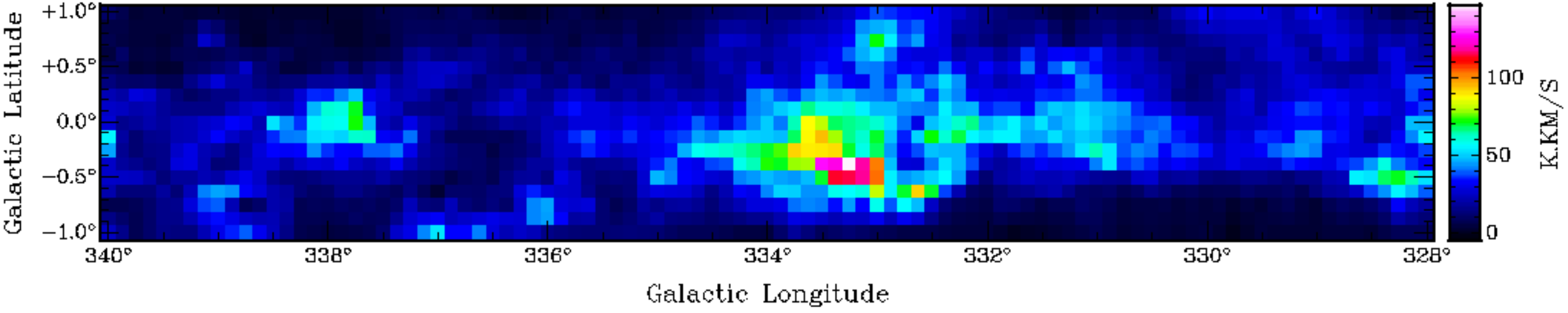}

\vspace{-8.7mm}\hspace{0.6mm}
\includegraphics[angle=0,scale=0.637]{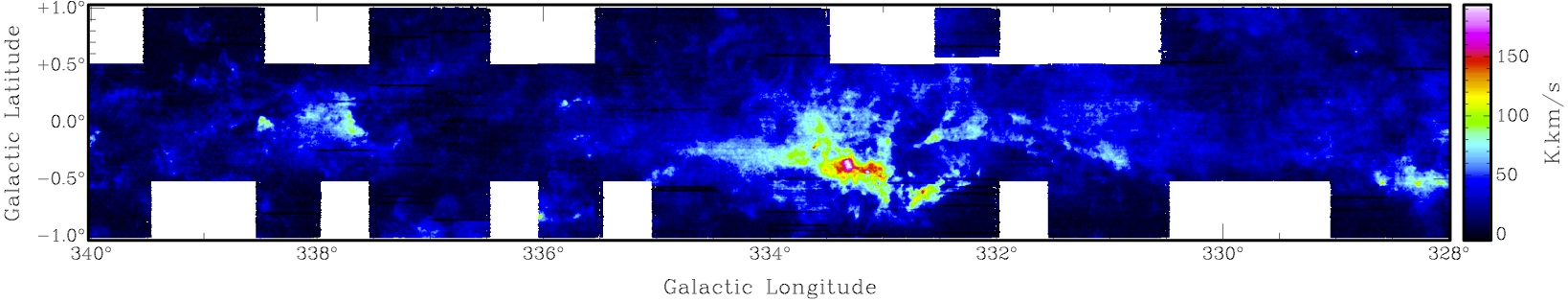}

\vspace{-7mm}\includegraphics[angle=0,scale=0.65]{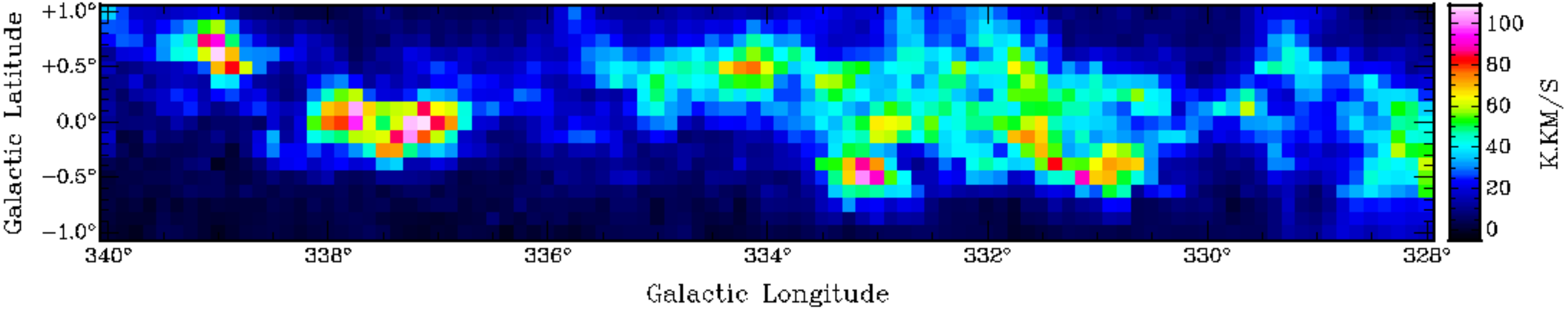}

\vspace{-8.7mm}\hspace{0.6mm}
\includegraphics[angle=0,scale=0.637]{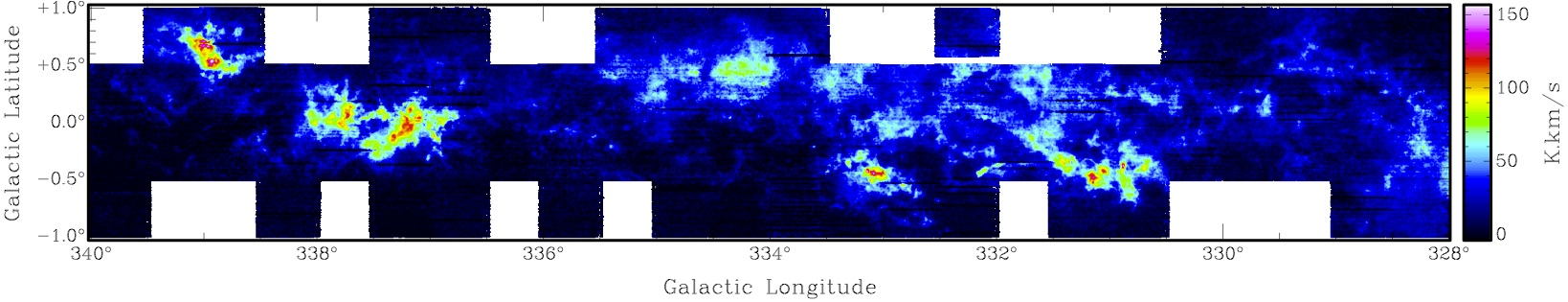}

{\color{white}
   \vspace{-152.5mm}
   \hspace{130mm}\parbox{46mm}{\bf GLIMPSE/MIPSGAL}

   \vspace{35mm}\hspace{13mm}\parbox{49mm}{\bf CfA \tco}

   \vspace{38.5mm}\hspace{13mm}\parbox{49mm}{\bf ThrUMMS \tco}
   
   \vspace{31mm}
   \hspace{13mm}\parbox{49mm}{\bf CfA \tco}

   \vspace{18mm}\hspace{13mm}\parbox{49mm}{\bf ThrUMMS \tco}
   
   \vspace{-100.5mm}\hspace{125mm}\parbox{46mm}{\bf $V_{LSR}$ = --55 to --40\kms}
   
   \vspace{31mm}\hspace{125mm}\parbox{46mm}{\bf $V_{LSR}$ = --55 to --40\kms}
   
   \vspace{40mm}\hspace{125mm}\parbox{46mm}{\bf $V_{LSR}$ = --75 to --55\kms}
   
   \vspace{9mm}\hspace{125mm}\parbox{46mm}{\bf $V_{LSR}$ = --75 to --55\kms}
} 
\vspace{23mm}
\caption{Comparison images of a 12\degree\ section of the Milky Way's 4th quadrant (4Q), as labelled. At 1\fmin2, ThrUMMS' high areal resolution allows us to explore the detailed physics of individual clouds and star-forming regions in much greater detail than before.  Simultaneously, we retain the ability to address global Galactic structure topics with a wide-field, uniform, sensitive, multi-transition survey.  In these pipeline-reduced ThrUMMS Data Release 3 (DR3) images, the rms noise is $\sim$2--3\,K\kms, and we have largely removed the mapping artifacts present in DR2 and earlier data.  Here we compare DR3 maps with equivalent Columbia-CfA survey \citep{dht01} data for two different velocity integrations across the 12\degree\ area.  The upper panels (velocity range --55 to --40\,\kms) are compared numerically in Fig.\,\ref{cfacal}.  The higher resolution of ThrUMMS maps compared to the CfA maps allows easy separation of individual continuum features that reside at different velocities (and hence at different distances).
}
\label{comps}
\end{figure*}

In \textsc{Livedata}, raw data for each observed field are bandpass-corrected and scaled by the measured \tsys\ to a relative antenna temperature scale $T_A^*$, using the ``reference'' method (with OFF positions as described in \S\ref{obsprocs}).  A linear baseline from a robust fit to signal-free channels (not including 100 channels at each end of each 4096-channel spectrum) is also subtracted, the frequency axis is rescaled for all topocentric to \vlsr\ corrections, and the result for each MOPS IF-zoom spectrum is written to a separate SDFITS file.  

\notetoeditor{}
\begin{figure*}[t]
\includegraphics[angle=0,scale=1.1]{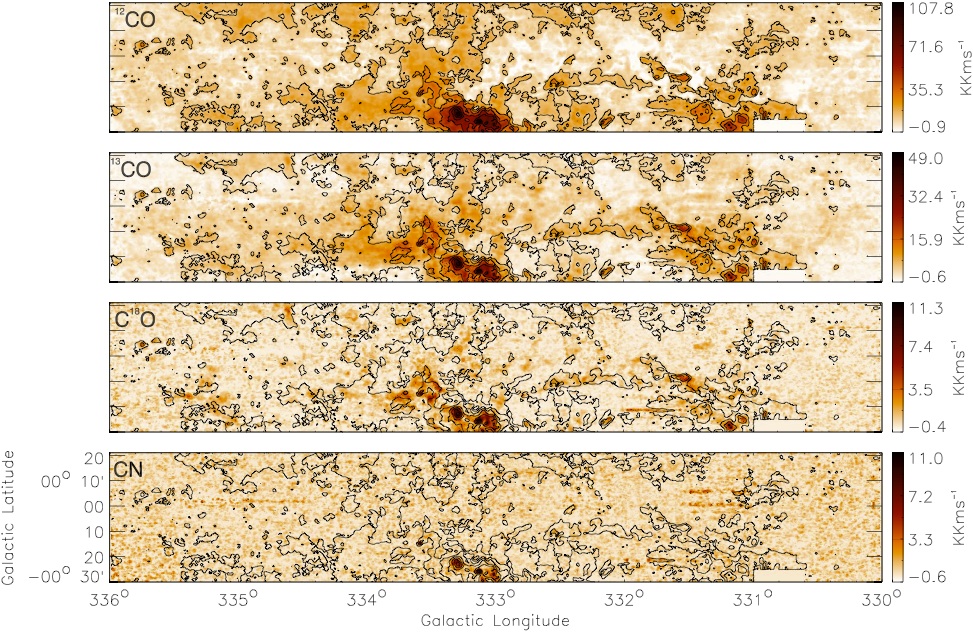}
\caption{Comparison of simultaneously-mapped ThrUMMS species on the $T_{A}^{*}$ scale, as indicated, integrated over --100\kms\ $<$ $V_{\rm LSR}$ $<$ +50\kms. The contours in each image are the same, of \tco\ at levels of 1, 2, ..., 7, and 8 times 13.6\,K\kms\ $\approx$ 20$\sigma$.  These DR3 maps cover half the area shown in Figure \ref{comps}. Note the different line ratios of various peaks, and that both \ceto\ and CN are easily detectable in several places. The \tco\ and \ttco\ emission is quite filamentary, similar to Hi-GAL images.
}
\label{dr3comps}
\end{figure*}

Most of our custom augmentations to this pipeline are effected between the \textsc{Livedata} and \textsc{Gridzilla} stages.  First, absolute flux calibration is performed on the SDFITS files produced by \textsc{Livedata}, as illustrated in Figure \ref{cals}.  Next, automated removal of known observing or processing artifacts is performed.  For example, in DR2 and earlier data releases, some artifacts (failed bandpass corrections for single 0.256-sec samples, bad calibrations due to observing malfunctions, or bad-weather ``stripes'' where baseline subtraction becomes less reliable) were not corrected, and were manifest in the data products.  While the fraction of data so affected is small, the effects on the cubes and maps can be quite large, since the individual errors can be of a large magnitude.  Left uncorrected, such data produce noticeable deleterious effects in the maps, such as ``bright'' or ``dark'' spots, strong ``edges,'' warped baselines, and so on.  However, these errors were easy to identify 
and then either flag or correct.  Thus in DR3, we have completely removed the single-pixel artifacts, systematic calibration errors, and some of the weather stripes, by new, custom, semi-automated routines  
written as a combination of IDL and Unix c-shell scripts. 
These additional calibration and editing steps improved the overall data quality to the point where the resulting noise levels were always consistent with theoretical expectations, given the observing conditions.

From this point the data processing resumes with the normal operation of \textsc{Gridzilla}, but with mapping parameters that are guided by the scientific motivation of defining the Galactic CO and CN emission to arcminute and sub-km\,s$^{-1}$ resolutions, given the unique features of the BSM technique.  During the observations, one integration cycle occurs approximately every half-arcminute interval, in both the scanning and scan-step direction (i.e., $l$ then $b$).  This is comparable to the native resolution of the Mopra telescope at 115 GHz \cite[approximately 33$''$;][]{L05}, so our mapping strategy under-samples that scale.  To form images with an effective resolution that {\em is} well-sampled, we define gridding parameters in \textsc{Gridzilla} that are $\sim$twice as large as those for normal OTF maps.  Thus, the map pixels are of size 24$''$ and the gaussian convolving beam is 68$''$, truncated within a 36$''$ gridding kernel, giving an effective resolution of 72$''$ in the maps, with a typical contribution of 40$\times$0.256-sec samples per pixel to the final spectra.  Therefore, the effective integration time is about 10\,s\,pixel$^{-1}$, and the output from \textsc{Gridzilla} preserves the intrinsic velocity resolution of MOPS, as listed in the footnote to Table \ref{linepars}, with the typical noise level also given there.  The final spatial sampling compares well with the typical spacing of $\sim$20$''$--25$''$ ($\approx$ 1 pixel) between raster lines in our VFM OTF maps.  Our mapping and data reduction procedures therefore return a product that is better than Nyquist-sampled at 72$''$ resolution, and cover an area of the Galaxy not feasible otherwise.

\notetoeditor{}
\begin{figure*}[t]
\vspace*{0mm}
\includegraphics[angle=0,scale=0.49]{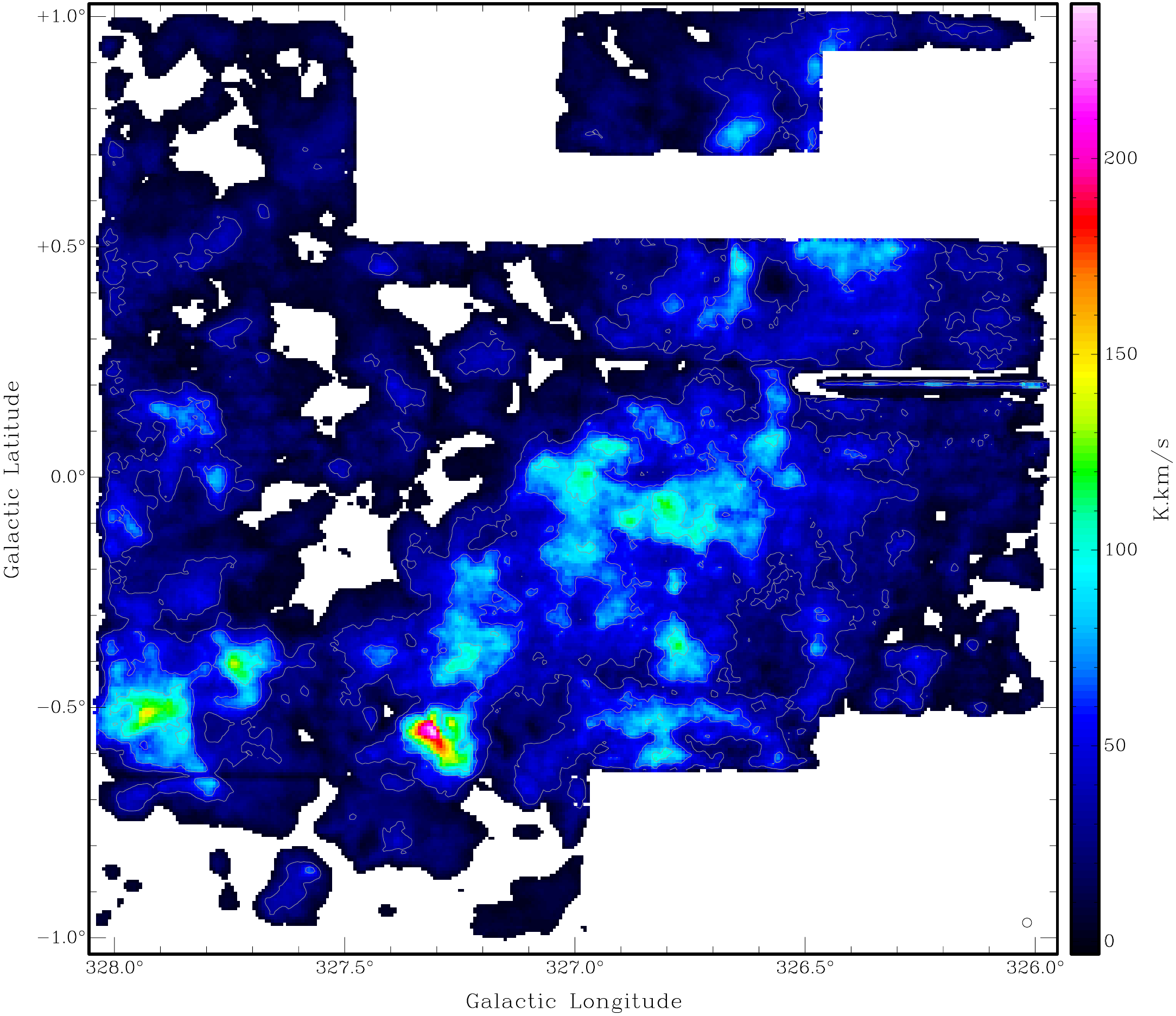}

\vspace{-6.9mm}\includegraphics[angle=0,scale=0.49]{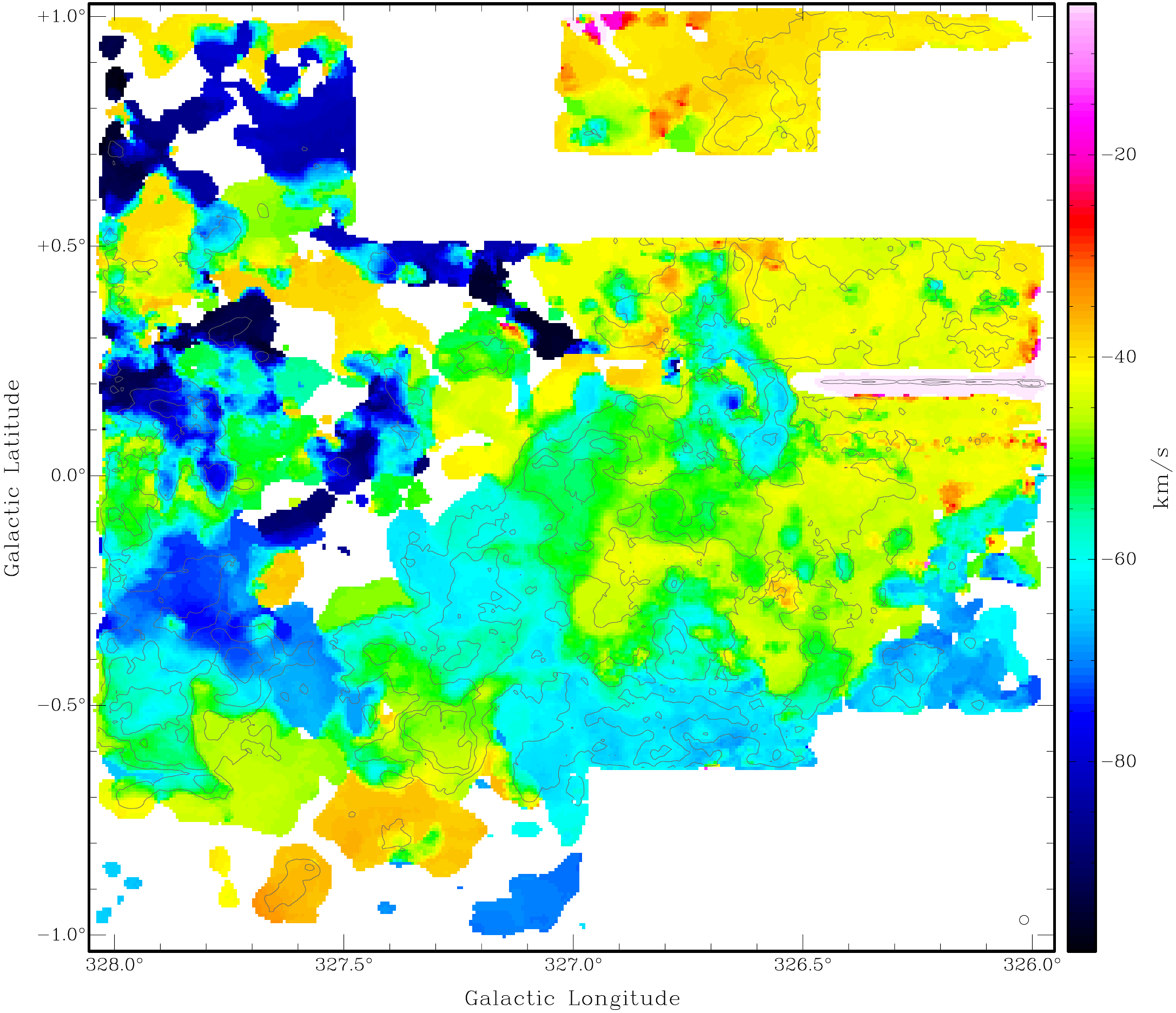}

\vspace{-168mm}\hspace{108.6mm}
\includegraphics[angle=0,scale=0.38]{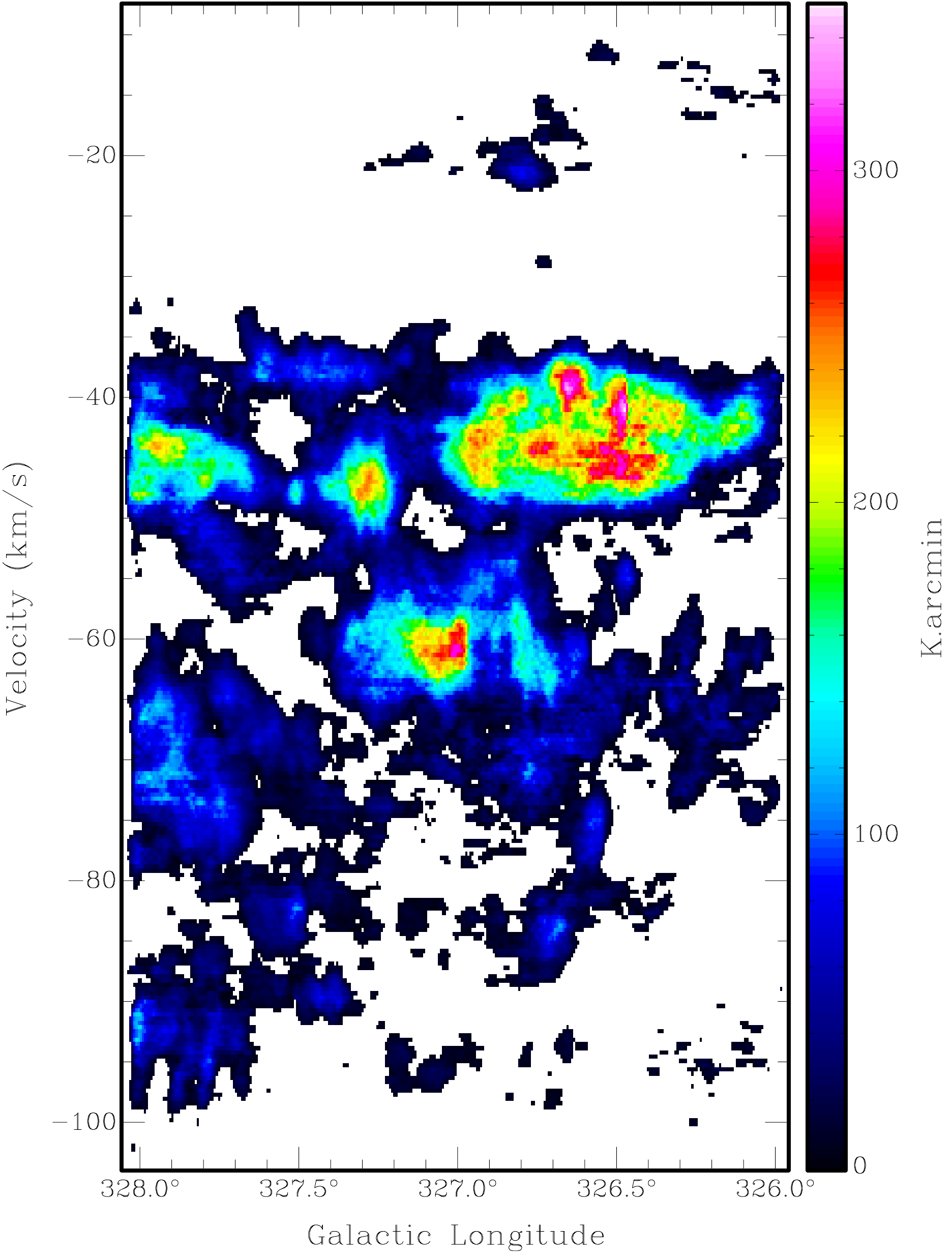}
\vspace{1mm}

\hspace*{112mm}\parbox{68mm}{
\caption{Sample downstream products from the ThrUMMS DR3 cubes, where a smooth-and-mask (SAM) algorithm has been applied using a custom IDL-\textsc{Miriad} shellscript to calculate moment maps.  (top-left) \tco\ intensity, integrated over all unmasked voxels from --150 to +50\,\kms; contours are at 10, 30, and 50 K\,\kms; (bottom-left) \tco\ velocity field, with contours as in the intensity map and evaluated over the same voxels and velocity range; and (right) \tco\ longitude-velocity diagram, integrated over all unmasked voxels in $|b|<1^{\circ}$.  Although artifacts due to poor weather are still visible in these data, these too will be removed in future DRs.  Note in particular that beam-sized noise modulation in the velocity field maps, common in such higher-moment calculations, are virtually eliminated in these DR3 products.  This is due to the superior filtering properties of the SAM method.  The effective Mopra beam is shown in the bottom-right corner of each of the intensity and velocity maps.
\label{depoxed}}}
\vspace{6mm}
\end{figure*}

Table \ref{linepars} also summarises some parameters of the data cubes for the 4 spectral lines we are mapping.  The per channel rms noise values drop with frequency since these lines are at the edge of the atmospheric 3mm band; thus, the \tsys\ for \ttco\ and \ceto\ are almost a factor of 2 less than for the \tco\ data.  Each species' cube is produced on the $T_A^*$ scale, and must be converted to $T_r^*$=$T_{\rm mb}$ for physical analyses.  Table \ref{linepars} gives both the main beam ($\eta_b$) and inner error beam ($\eta_c$) efficiencies we recommend for each spectral line, based on the measurements of \citet{L05}.  These efficiencies are respectively appropriate \citep{ku81} for compact sources of $\sim$one intrinsic Mopra beam or less, and for slightly more extended (a $\sim$few arcmin) sources.  For most emission seen in the ThrUMMS maps, $\eta_c$ will be more often used, and the DR cubes are all scaled to $T_r^*$ by $\eta_c$ (although see the discussion in \S\ref{cfa-comps} about refinements to this scaling).

The next step in our augmented processing pipeline after the \textsc{Gridzilla} stage is to calculate moment maps for the line emission.  We describe these methods in \S\ref{sam}.

\subsection{Verification\label{cfa-comps}}

The final step in the data reduction pipeline is to check that we have constructed maps that are at least consistent with previous surveys, e.g., the CfA data of \citet{dht01}.  As can be seen in Figure \ref{cfacal}, the ThrUMMS data do indeed compare well: averaging over a variety of regions, we derive an overall scale factor between the two surveys of  $T_r^*$(ThrUMMS) = (0.90$\pm$0.05) $\times$ $T_r^*$(CfA), which is within the overall calibration uncertainties given in Figure \ref{cals}.  We consider this result (close to unity, with a small variance) a strong verification of the quality-control procedures described above.

\notetoeditor{}
\begin{figure*}[ht]
\includegraphics[angle=0,scale=0.21]{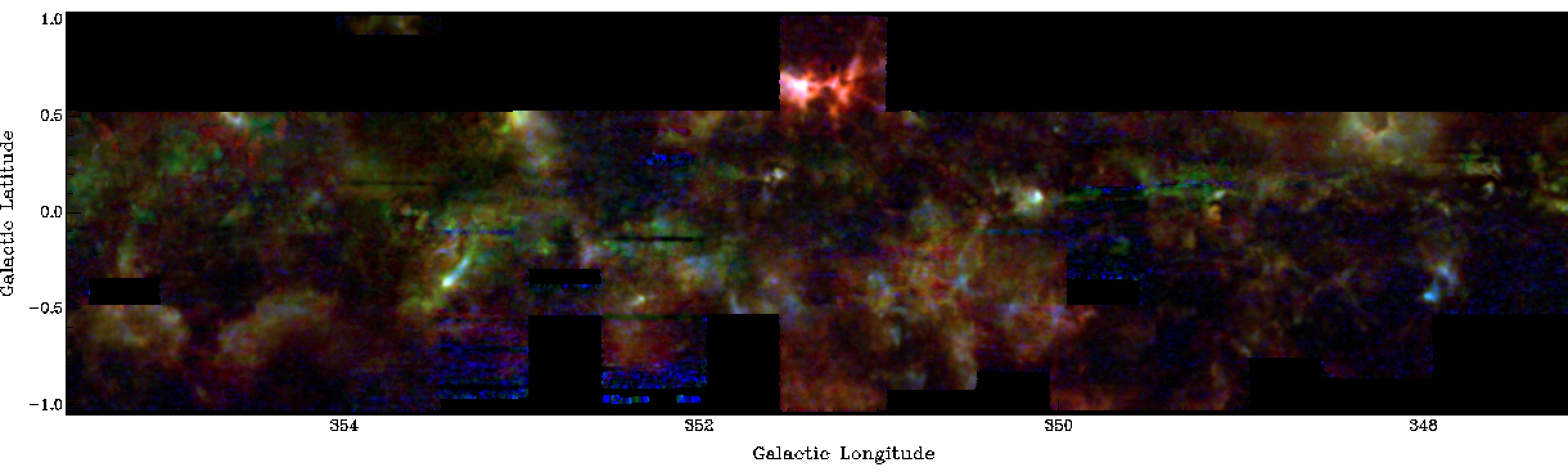}
\caption{Mosaic of DR3 integrated intensities (zeroth moment) for the three CO isotopologues across $\sim$9\degree\ of longitude within the ThrUMMS survey area.  Here, red = \tco, green = \ttco, and blue = \ceto.  Note the wide variation in line ratios across the Galactic Plane, indicated by the varying colors in this image.  E.g., the prominent red cloud at $l\sim$351\fdeg5 is relatively bright in \tco, but relatively faint in \ttco\ and \ceto.  In contrast, there are numerous small ``blue'' features throughout the image, where \ceto\ is relatively bright, but \ttco\ and \tco\ less so.
}
\label{colour}
\end{figure*}

Even so, the conversion factor is slightly but systematically below 1.  This may be due to two effects.  One is that the effective coupling between the sky emission and the inner error beam ($\sim5'$) of the Mopra telescope is not as uniform as is implicitly assumed by the use of $\eta_c$ = 0.55--0.57 in Table \ref{linepars}.  The conversion suggests instead that an efficiency of 0.50 may be better value, i.e., about 10\% below $\eta_c$, but 15\% above $\eta_b$.  This would be true if the sky emission only partly fills (perhaps to $\sim$60\% of) the inner error beam, giving an effective $\eta$ between these nominal values.  However, the appearance of the maps (especially the \tco\ from which this result is obtained) is not completely consistent with this idea: over large areas the \tco\ cloud structure is very extended, although it is compact in other areas, so we might expect a more variable conversion between the ThrUMMS and CfA brightness scales, rather than the fairly systematic trend evident in Figure \ref{cfacal}.

An additional factor may be that our BSM technique is systematically missing a few percent of the line flux from all mapped areas, due to the undersampling of the emission structure (see \S\ref{BSM}).  That is, between raster lines there may be small, bright, compact (i.e., beam-sized) structures whose total flux would not be fully measured by the beam-sampling, yielding an apparently lower brightness scale for the maps after interpolating between the raster lines.  Another way to understand this issue is by analogy with interferometry: there are scales to which our maps are less sensitive than if they were properly Nyquist-sampled.  Tests of this effect in \tco\ maps of the CHaMP areas (which {\em are} Nyquist-sampled; Barnes et al, in prep.) suggest that it may be relevant, but we defer that discussion to that work.  For the purposes of this paper, we use these explanations as working assumptions to account for the 0.90 conversion factor.  Thus, while we quote all results herein on the $T_r^*$ scale as derived with the tabulated $\eta_c$ factors, the reader should be aware that these brightness levels may need to be increased by a further $\sim$10\%.

\section{Data Products}\label{results}
\subsection{Online Access of Spectral Line Cubes\label{access}}

From its inception, ThrUMMS was designed to be an open-access project, with frequent Data Releases as progress is made in the observing and data reduction.  As of 2014 May, we have placed a large amount of data on the website http://www.astro.ufl.edu/thrumms, freely available to the community.  For the \tco\ and \ttco\ cubes, the data are at DR3 level and organised into 6\degree-wide ``Sectors'' (i.e., each including up to 48 contiguous half-degree fields; see Figs.\,\ref{comps}--\ref{lv}).  For the \ceto\ and CN data, we currently have available the DR2 cubes, organised into 2\degree-wide blocks; DR4 is planned to include the augmented artifact-correction for the latter two species, as well as wider area coverage from the 2014 observing season, and is anticipated to be available in mid-to-late 2015. 
At the present time, these data cubes cover $\sim$80\% of the 60\degree$\times$2\degree\ target area for all 4 species.  With the anticipated completion of observing in 2015, later Data Releases will continue to be provided as processing continues and further data products can be assembled.

\subsection{Moment Maps and Initial Processing\label{moments}}

We use the \textsc{Miriad} package \citep{s95} to process the data cubes into moment maps and LV diagrams, and perform other basic analyses, with visualisation provided by the \textsc{Karma} package \citep{g97}.  Figure \ref{comps} gives some sample comparisons between ThrUMMS maps and other surveys, while Figure \ref{dr3comps} shows sample VFM data for our 4 molecular transitions.  Note the high S/N of the \tco\ and \ttco\ maps, and even though the \ceto\ and CN maps have noticeably smaller S/N, the peaks in these maps (around the positions of the brightest \tco\ or \ttco\ emission) are still detected with good reliability (S/N \gapp\ 5--10).  In Figure \ref{depoxed} 
we show further examples of higher moment and LV maps in the \tco\ line, demonstrating the very high image fidelity achieved across extended emission regions.  

\subsection{DR3 and Smooth and Mask\label{sam}}

The example moments maps of Figure \ref{depoxed}  were processed with a {\em smooth-and-mask} (SAM) algorithm in order to improve the fidelity of these products.  As part of this algorithm, the DR3 cubes are binned by 4 channels in velocity (see Table \ref{linepars}), which improves the S/N in the public data cubes and reduces the file sizes.  (The full-resolution cubes are, however, still available on request.)  Our SAM algorithm is similar to those described by \citet{rbv90}, \citet{dht01}, or \citet{d11}.  We form a data cube that has been smoothed/convolved in all three dimensions by a factor of 2 (i.e., beyond the 4-channel binning of the public data), a 5$\sigma_{\rm rms}$ threshold was calculated, and voxels (where a voxel is defined here as a single ($l$,$b$,$V$) element) below this level in the smoothed cube were used to define a blanking mask for voxels in the {\em unsmoothed} data cube.  Moments are then calculated only with the unmasked voxels.

\notetoeditor{}
\begin{figure*}[ht]
\includegraphics[angle=-90,scale=0.65]{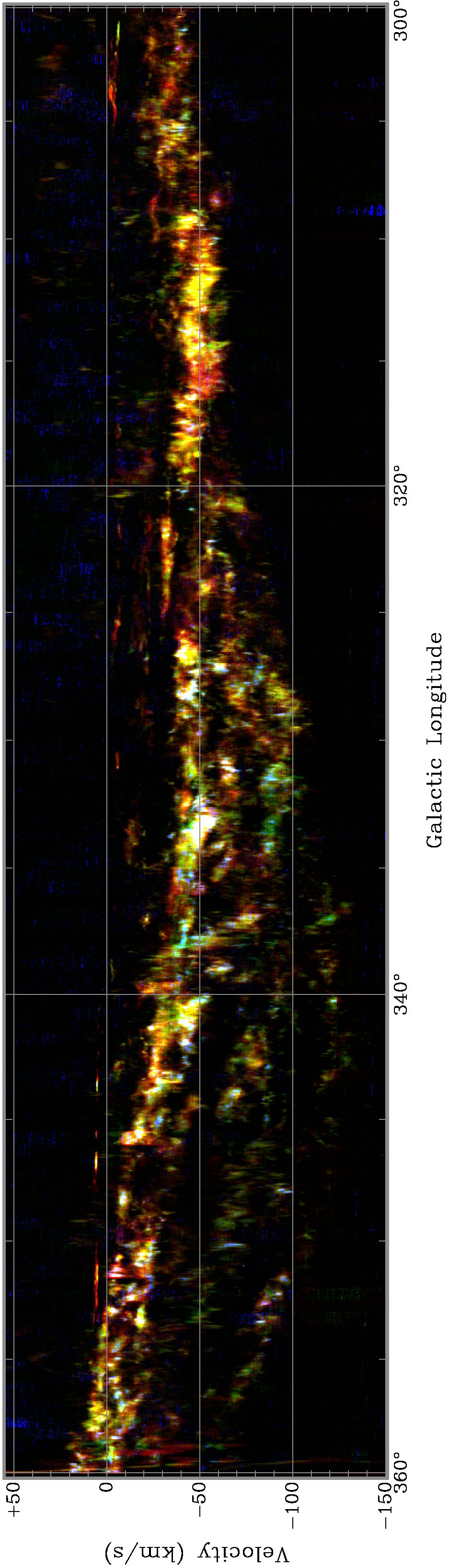}

\vspace{0mm}
\caption{Mosaic of ThrUMMS pre-DR4 $l$--$V$ images (SAM-based zeroth moment integrated across all $b$) for \tco\ (red), \ttco\ (green), and \ceto\ (blue) across the 4Q.  As in Fig.\,\ref{colour}, we see wide variations in line ratios across the Galactic Plane from the varying colours in this image.  The data maximum, saturated colour, and 3$\sigma$ (almost black) levels in this image are respectively at 577, 248, and 12\,K\,arcmin (\tco), 187, 63, and 5\,K\,arcmin (\ttco), and 37, 5, and 3\,K\,arcmin (\ceto).
}
\label{lv}
\vspace{0mm}
\end{figure*}

Construction of moment maps, however, depends strongly on the particular investigation one wishes to pursue with the data cubes.  There is a large range of input parameter combinations possible for such maps, including angular extent, velocity range, projections, SAM parameters, uncertainty estimation, etc.  For this reason, moment maps using the above techniques and based on the public data cubes are not currently part of the DRs, but are available upon request (such requests are processed on a best-efforts basis).

We do plan, however, to include sets of standardised moment maps in future DRs, once we have re-observed most of the weather-damaged areas and made the data cubes' noise statistics more uniform.  Meanwhile, users can use many standard astronomical or image processing packages to perform their own analyses, including complex sorting and assessment algorithms such as Principal Component Analysis \citep{hs97,L09}, ISM structural analyses like Velocity Component Analysis \citep{LP00}, or Spectral Correlation Functions \citep{rgw99}.

\subsection{The Full Survey\label{full}}
The full survey data cubes provide an extremely rich dataset for current and future investigations.  This richness means it is difficult to render, whether on the printed page or a web browser, the detailed physical insights contained in the data.  Nevertheless, the reader may gain some intuitive feeling for this richness by means of colour-compositing the different species' integrated intensity maps.  We show examples of this in Figure \ref{colour}, combining the DR3 \tco, \ttco, and \ceto\ integrated intensities, and in Figure \ref{lv} for pre-DR4 versions of the \tco, \ttco, and \ceto\ data, each showing clear positional variations in the line ratios as variations in the rendered colours.

Similarly, compressed image versions of the spectral line integrated intensities across the full survey area, in all four species, are given in the Appendix.  The reader should note, however, that the Appendix images, and the data used in the analysis presented below, have been manually edited from the DR3 versions available on the ThrUMMS web site to exclude most of the artifacts generated during poorer observing conditions (``weather stripes'').  These excisions are manifest in the Appendix figures as half-degree-wide gaps in certain fields; the gaps themselves are not the weather stripes.  These blanked areas cover only $\sim$2\% of the mapped area, however.

\section{First Science Results}\label{disc}

ThrUMMS is designed to provide legacy data for a wide array of science investigations, with specifically-focused applications in astrochemistry and the physical state of GMCs; the neutral/molecular ISM connection; assessment of the turbulence and dynamical parameter space on both a targeted (i.e., specific regions) and holistic level; 
and Galactic structure and kinematics.  But the quality and scope of the dataset is also suitable for studies where our attention is not currently focused.  Because ThrUMMS is an open project, we welcome new initiatives from members of the community not yet involved with ThrUMMS, and indeed encourage those who are so inclined to collaborate with us in developing new software or data products to expand ThrUMMS' impact beyond our original vision.  

Our first detailed analysis of a specific region, including derivation of kinematic distances, the holistic properties of the GMC population, their relationship to the HI and cold dust (Hi-GAL+MIPSGAL) emission, and implications for the complexes' inferred star formation rate, is given in Paper II of this series \citep{q15} for the G333/RCW\,106 complex.  In contrast, we describe here some first results of the global analysis of the CO-isotopologue lines across the 4Q.

\notetoeditor{}
\begin{figure*}[ht]
\hspace{-1mm}\includegraphics[angle=0,scale=0.17]{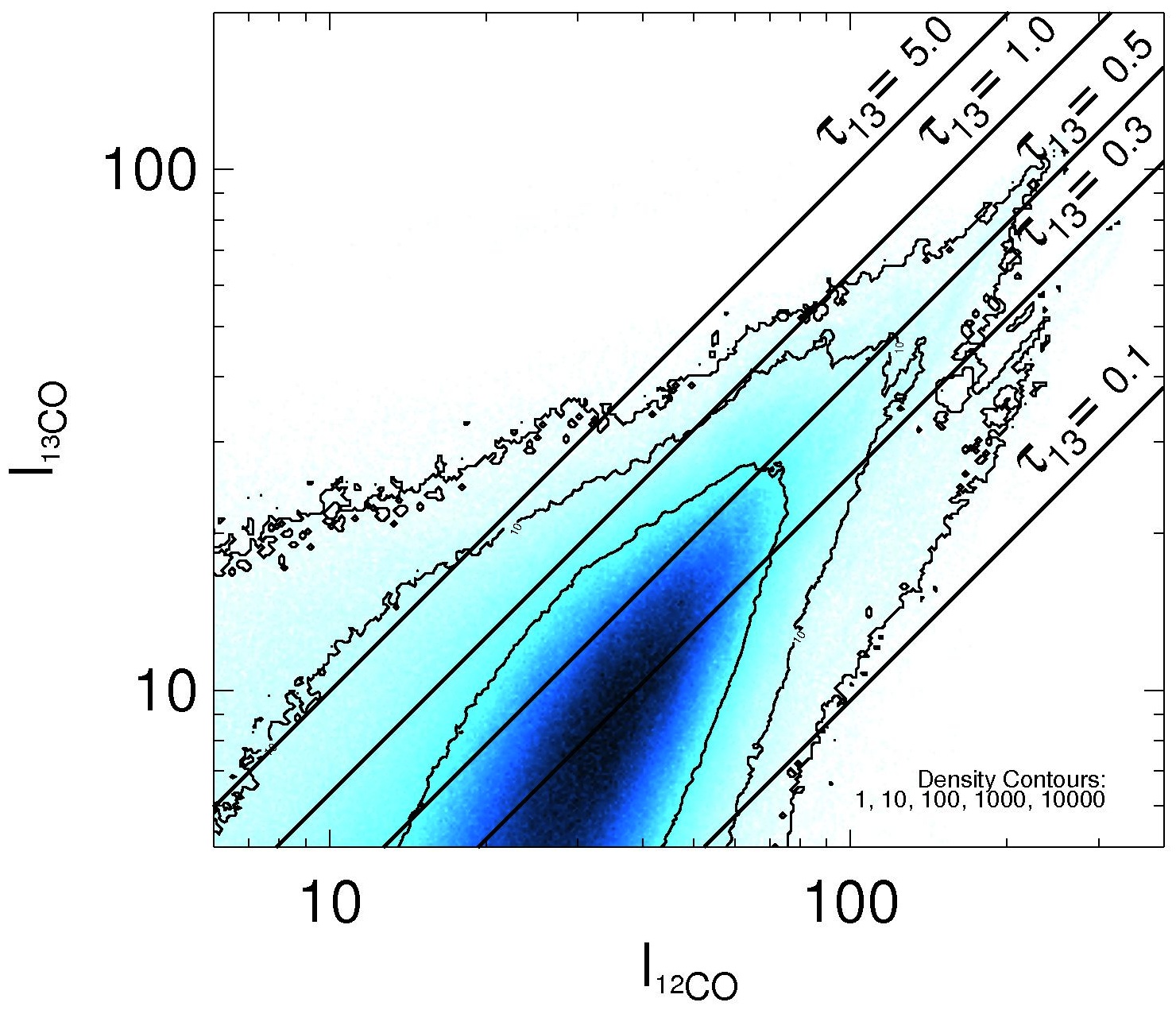}

\vspace{-76.6mm}
\hspace{88mm}\includegraphics[angle=0,scale=0.62]{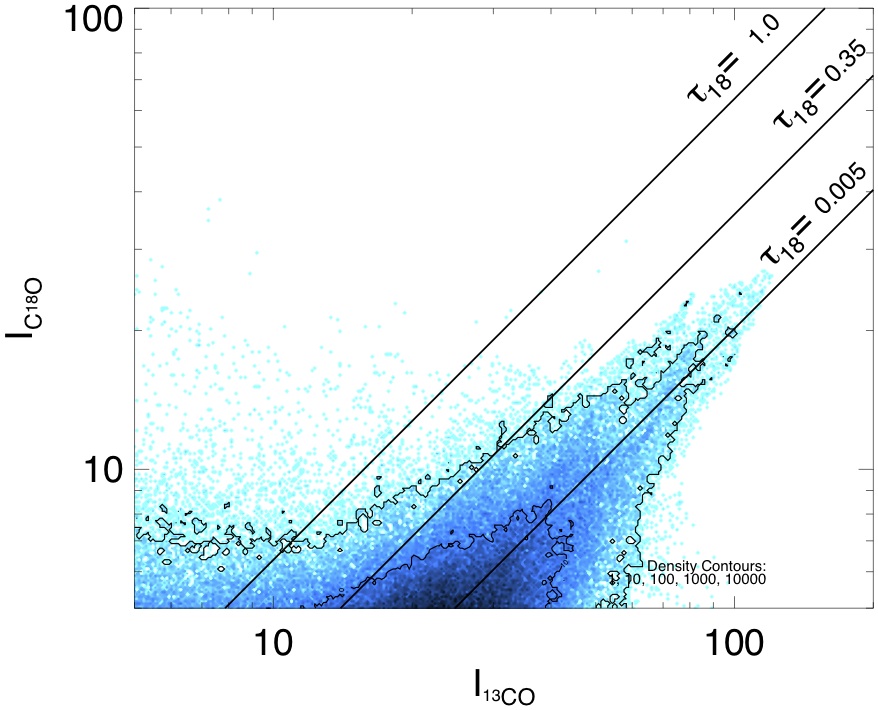}

\vspace{1mm}
\caption{({\em left})  Plot of voxel density (coded by color and contoured on an arbitrary scale) in the $I_{\rm ^{13}CO}$--$I_{\rm ^{12}CO}$ plane, where the intensity is integrated over 1\kms\ velocity bins, from all-4Q mosaics similar to Fig.\,\ref{colour}.  The lines are labelled by $\tau_{\rm 13}$ for an assumed $R_{\rm 13}$ = 60.  ({\em right})  Similar plot for \ceto\ and \ttco.  The lines are labelled by $\tau_{\rm 18}$ for an assumed $R_{\rm 18}$ = 10.  See text for further discussion.
}
\label{ratios34}
\end{figure*}

\subsection{Optical Depth and Excitation Temperature Variations}\label{tauratios}

With our comprehensive coverage of a large portion of the Milky Way in multiple species, we have an unprecedented opportunity to use the molecular lines to investigate global GMC conditions.  We examine now the variation in the CO line ratios (and consequently, the variation in the physical conditions in GMCs) with position in the Galaxy, as shown visually in Figures \ref{colour} and \ref{lv}.  This rendering can be intuitively understood in terms of radiative transfer physics in the following way.

The clouds rendered in redder colours show where the \tco\ is relatively much brighter {\em on average} than either \ttco\ or \ceto, i.e., where the \tco/\ttco\ and \tco/\ceto\ brightness ratios are near their maximal values.  This must therefore indicate areas of relatively low optical depth, since the intrinsic abundance ratios are also quite large (any line saturation = high optical depth would produce brightness ratios significantly less than the respective abundance ratios).  At the same time, the \tco\ lines in such areas are quite bright, typically with $T_r^*$ \gapp\ 30\,K.  So the ``red'' clouds are typically those of low $\tau$ and high $T_{\rm ex}$, or ``warm and translucent'' clouds.

In contrast, green or even cyan/blue areas are where either \ttco, \ceto, or both are much brighter compared to \tco\ {\em than on average}.  This is only likely to occur where the optical depth is high enough to produce some level of saturation in either line, so that the \tco/\ttco\ and \tco/\ceto\ brightness ratios are closer to unity than average.  At the same time, the \tco\ emission in such areas is not particularly bright; coupled with the higher optical depth, this means that the excitation temperature in such clouds is likely to be somewhat low.  So the ``green'' and ``blue'' clouds will typically represent areas of high $\tau$ and low $T_{\rm ex}$, or ``cold and opaque'' clouds.

Therefore, we understand that the line ratios provide a simple yet powerful way to derive basic physical conditions in Galactic GMCs from an elementary radiative transfer analysis.  The rendered colours make these relationships intuitively clear: they are set so ``white'' or ``neutral'' colours represent the {\em average} line ratios and {\em not} where the line ratios are close to 1.

Other workers have used a similar approach to derive cloud properties from CO line ratios; examples include \citet{mlb83}, \citet{wlb08}, and \citet{h13}.  Each of these, however, uses some simplifying assumptions about the excitation temperature, abundance ratios, or other radiative transfer parameters, and typically for relatively small samples of clouds covering at most 10s of pc.  Ours is the first such study of which we are aware that attempts this analysis uniformly across many kpc of the Galaxy; at the same time, we retain the pc-scale resolution necessary to characterise the clumps that give rise to individual star clusters.

In the analysis that follows, we use the full 3D iso-CO data cubes, but binned in velocity to 1\kms\ channels, to compute all physical quantities.  We do this in order to fully track the variations of \tex. $\tau$, and other parameters with position and velocity, but also to boost the S/N in the \ceto\ data especially.  In what follows, therefore, the notation $\int$d$V$ refers only to these 1\kms\ integrals, which are still voxels in the binned cubes, unless otherwise stated.

For abundance ratios $R_{\rm 13}$ = [\tco]/[\ttco] and $R_{\rm 18}$ = [\ttco]/[\ceto], the optical depths in the different isotopologues will follow $\tau_{\rm 12}$ = $R_{\rm 13}\tau_{\rm 13}$ and $\tau_{\rm 13}$ = $R_{\rm 18}\tau_{\rm 18}$.  Then the radiative transfer equation for each spectral line (in the Rayleigh-Jeans limit, where $J_\nu$ $\propto$ $T$), 
\begin{equation} 
	T_{\rm mb} = (T_{\rm ex}-T_{\rm bg})(1-e^{-\tau})~,
\end{equation}
can be combined to convert the line ratios to an optical depth, with the further assumption of a common excitation temperature $T_{\rm ex}$.  In general, this last condition may not hold everywhere: the lower \ttco\ optical depth means that this species will more often be subthermally excited compared to \tco, which has a lower ``effective'' density due to strong radiative trapping \citep{e99,hd15}.  Nevertheless, this assumption will only act to underestimate the resultant \tco\ column density (see \S\ref{coldens}).  

\notetoeditor{}
\begin{figure*}[ht]
\hspace{0mm}
\includegraphics[angle=0,scale=0.14]{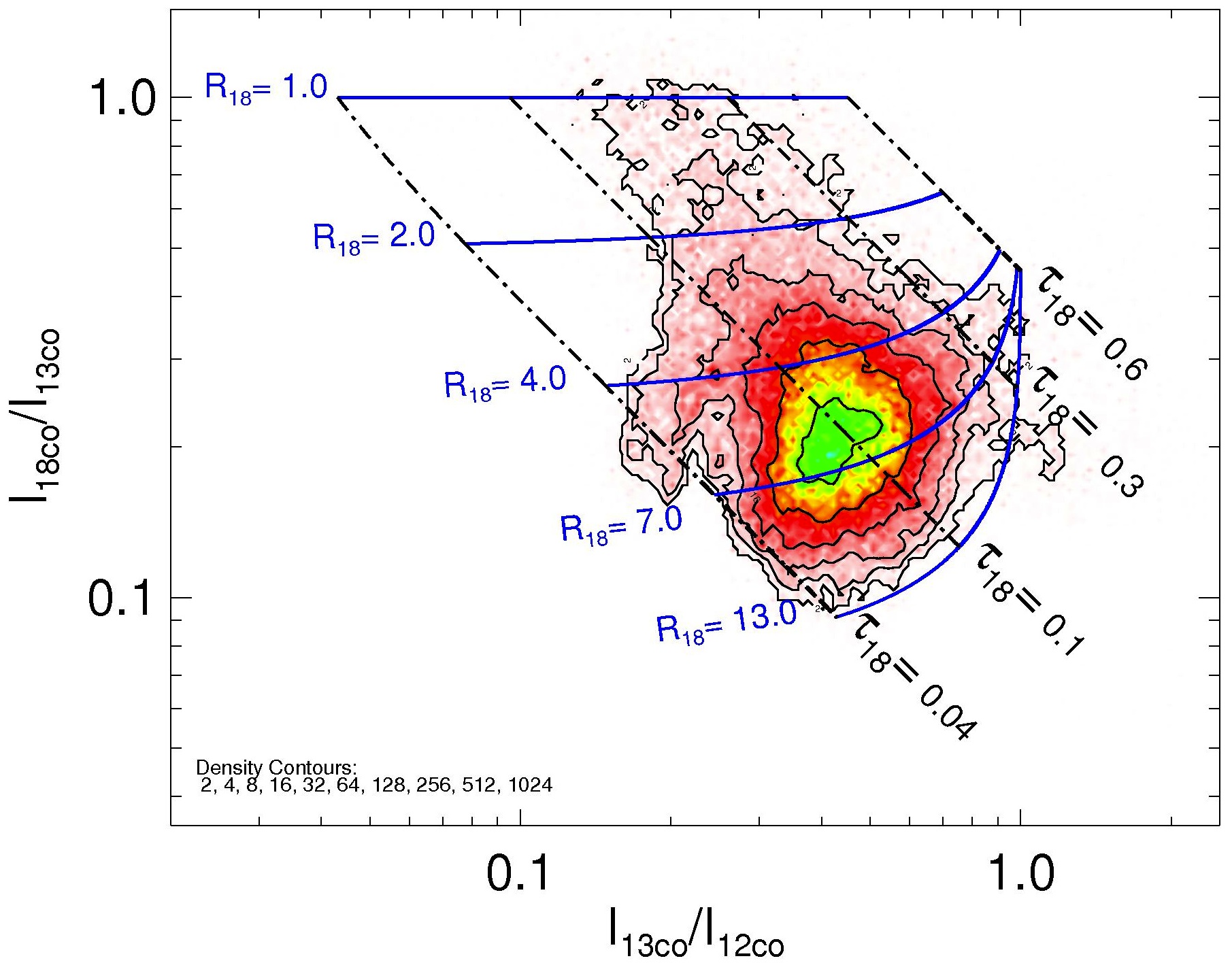}
\includegraphics[angle=0,scale=0.14]{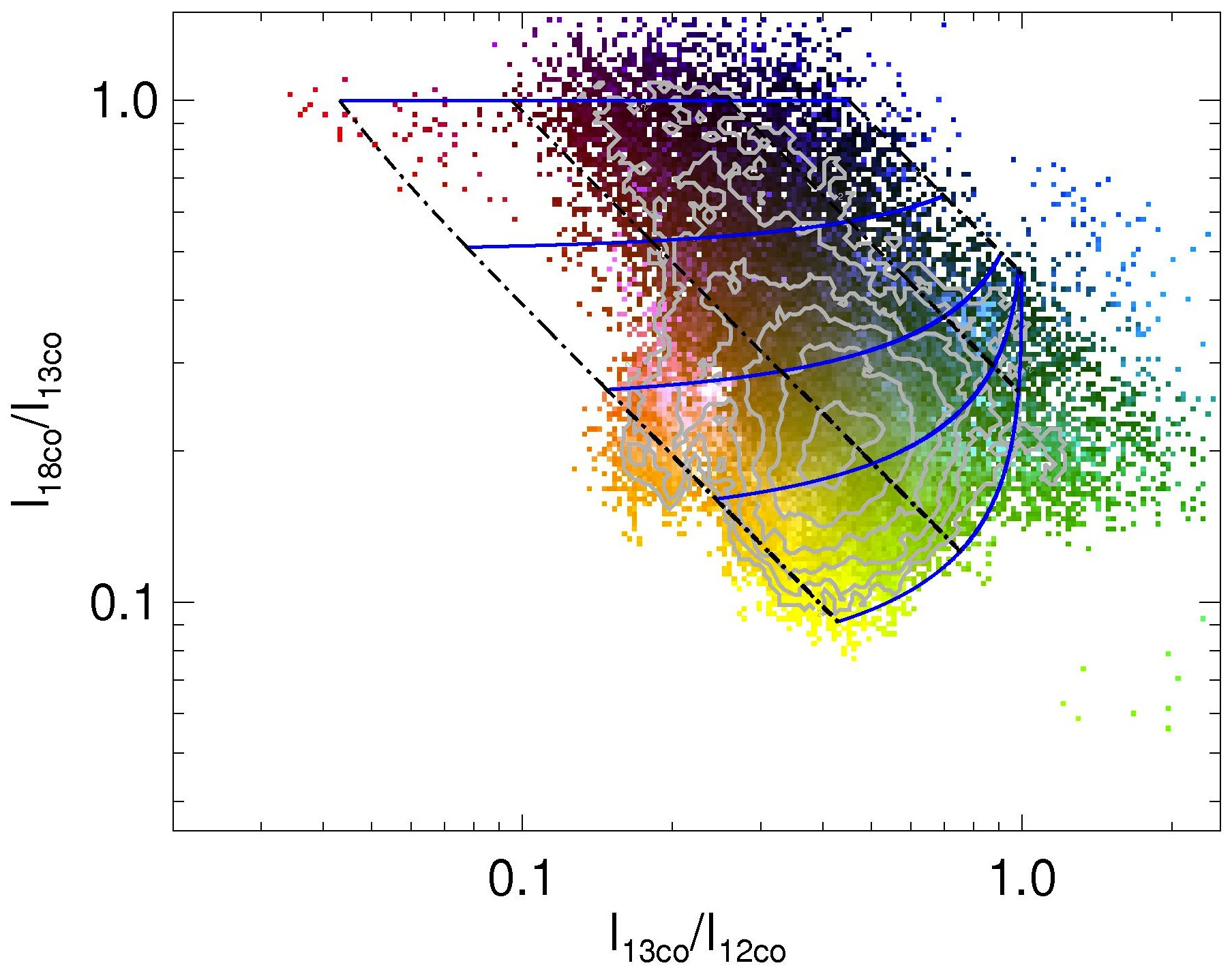}
\caption{Two plots of the same CO line ratio data.  ({\em left}) Voxel density (coded by colour and contoured on an arbitrary scale) in the CO line ratio plane.  Overlaid are simple radiative transfer models based on eqs.\,(2--3); see text for further discussion.  ({\em right}) The same CO line ratio plane as in the left panel, but the colours in this plot represent the approximate appearance of pixels in composite images like Figs.\,\ref{colour}, \ref{lv}, and in the Appendix.  The overlays are the same as in the left panel.
}
\label{ratios6}
\end{figure*}

With these assumptions, the brightness ratio in the ThrUMMS maps should obey
\begin{equation} 
	\frac{T_{\rm 13}}{T_{\rm 12}} = \frac{1-e^{-\tau_{\rm 13}}}{1-e^{-R_{\rm 13}\tau_{\rm 13}}}   ,
\end{equation}
and a similar equation applies for \ceto\ and \ttco.  For typical values of $R$ and $\tau$ (especially for $R_{\rm 13}$), the denominator in eq.\,(3) will often be close to unity.  Also, for any reasonable value of $R_{\rm 13}$, the \tco\ will be virtually opaque almost everywhere, and $T_{\rm 12}$ = $(T_{\rm ex}-T_{\rm bg})$.

The visual representations of line ratio variations, as seen in Figures \ref{colour} and \ref{lv}, can be quantified and understood in Figure \ref{ratios34}.  Here we see that, in the left panel, there is a very wide range of ratios log($I_{\rm ^{13}CO}$/$I_{\rm ^{12}CO}$) = log($\int T_{\rm ^{13}CO}$d$V$/$\int T_{\rm ^{12}CO}$d$V$), from $\sim$\ +0.2 in the upper-left of this panel (seen as green areas in Figs.\,\ref{colour} and \ref{lv}) to $\sim$\ --1.1 in the lower-right of Figure \ref{ratios34} (redder areas in Figs.\,\ref{colour} and \ref{lv}).  Similarly in the right panel, we see a range of ratios for log($I_{\rm C^{18}O}$/$I_{\rm ^{13}CO}$) = log($\int T_{\rm C^{18}O}$d$V$/$\int T_{\rm ^{13}CO}$d$V$), from $\sim$\ +0.5 in the upper-left of this panel (seen as blue or cyan areas in Fig.\,\ref{colour}) to $\sim$\ --1.2 in the lower-right of Figure \ref{ratios34} (green or yellow areas in Fig.\,\ref{colour}).

Now we can apply eq.\,(3) to the data in Figure \ref{ratios34}, where in each panel we show lines of constant $\tau$ for an assumed $R$. As expected, low optical depth for the line pairs occurs to the right of each panel, while high optical depth lies to the left.  Interestingly, in Figure \ref{ratios34} we see several groups of high-brightness emission peaks (i.e., the brightest clumps in Figs.\,\ref{colour}, \ref{lv}) as separate extensions of points, parallel to lines of constant $\tau$, in the distribution to the upper right of each panel.  These extensions apparently correspond to groups of bright clumps (complexes) with different regional $\tau$ and $T_{\rm ex}$ (see \S\ref{heavy}).

The plots in Figure \ref{ratios34} can be combined into a single ``CO-colour'' plot (Fig.\,\ref{ratios6}), where a voxel's position in this plot corresponds to its colour rendering in images like Figure \ref{colour} or \ref{lv}.  This can be compared with the analytical relations derived above (eqs.\,2--3) which are shown as overlaid lines in Figure \ref{ratios6}.  From this comparison it is immediately obvious that we cannot model the most extreme ``green'' and ``blue'' voxels with the assumptions used above (i.e., a common $T_{\rm ex}$ and fixed ratios $R$$>$1).  For this small number of voxels, the model would require either (1) ratios $R$$<$1 (e.g., due to strong fractionation or other selective effects), or (2) a lower $T_{\rm ex}$ in the more abundant and optically thick species (e.g., due to a colder foreground layer and warmer cloud centre).  Fractionation effects for both \ceto\ and especially \ttco\ \citep{g85} do occur, as does self-absorpton from colder cloud envelopes, so undoubtedly these both contribute to deviations from the simple radiative transfer picture presented here.  Note also that single clouds may be represented by a range of loci in Figure \ref{ratios6}.  So, while we can understand most of the line ratio variations in Figures \ref{colour} and \ref{lv}, i.e., the lower and leftmost boundaries of the distribution in Figure \ref{ratios6}, eventually a more sophisticated radiative transfer approach will be needed to fully understand all the line ratios, because the interplay between optical depth, excitation, and abundance is more complex than the above treatment allows.

\begin{figure*}[ht]
\centerline{\hspace{0mm}
\includegraphics[angle=0,scale=0.259]{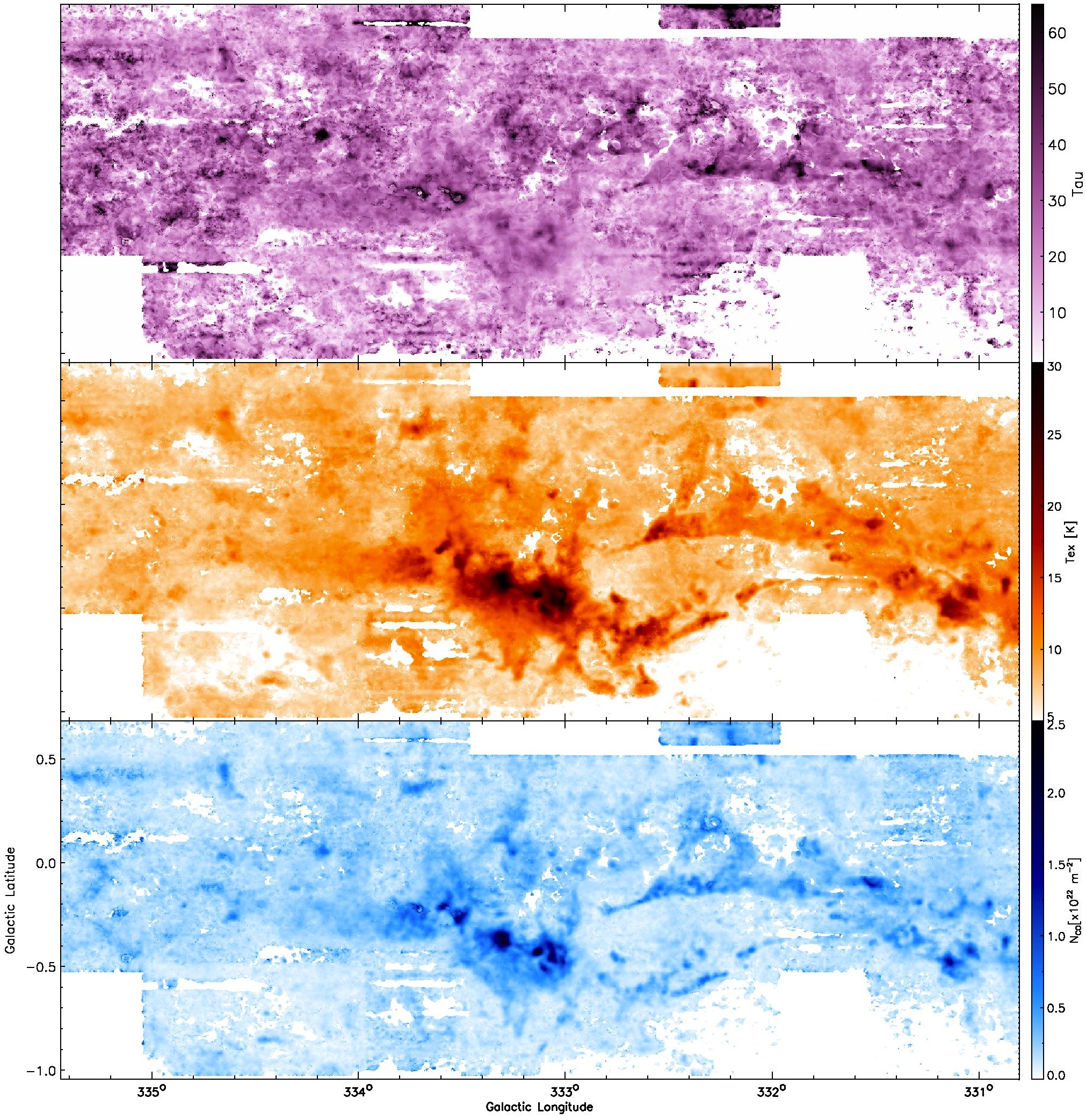} 
}
\caption{Sample integrated maps of $\tau_{\rm 12}$ ({\em top, purple}), $T_{\rm ex}$ ({\em middle, orange}), and $N_{\rm CO}$ ({\em bottom, blue}) near $l$ = 333\degree, obtained from the respective cubes of these quantities as described in the text.  Note in particular that the column density map is noticeably more filamentary than either the very clumpy $\tau$ map, the rather amorphous \tex\ map, or the $I_{\rm CO}$ (Fig.\,\ref{sector4}) map.
}
\label{tautex}
\end{figure*}

Nevertheless, we can use our treatment to estimate the molecular column density distribution in the Galactic Plane (\S\ref{coldens}).  We use the line ratios to evaluate $\tau$ via eq.\,(3) at each 1\kms-binned voxel, and then invert eq.\,(2) to also obtain $T_{\rm ex}$ at each voxel, producing all-4Q cubes of both parameters.  Finally, we form moment maps from each of these cubes in the normal way, e.g., integrating over large velocity ranges to obtain the total $\tau$, peak or average $T_{\rm ex}$, or total $N_{\rm CO}$ (see also \S\ref{coldens}).  We show examples of these results in Figure \ref{tautex}; although these are moment maps, the structures they exhibit are typical of the cubes over which they were integrated, although there are fewer visible complexities in any given velocity channel.

\notetoeditor{}

The $\tau$ panel in Figure \ref{tautex} (top) shows a very high-contrast image: all the highest optical depth structures are quite compact, typically having size scales of $\sim$1$'$--5$'$.  It may not be coincidental that this is also the size scale ($\sim$a few pc at 3\,kpc distances) of dense clumps, IRDCs, and clusters.  Lower optical depth structures are not quite so spatially compact, and are somewhat widely-distributed.

In contrast, the \tex\ panel of  Figure \ref{tautex} (middle) shows a dramatic difference: nearly all structures are relatively large-scale, and somewhat amorphous and/or low-contrast.  While the warmest structures are quite extended, at least 5$'$, moderate-temperature clouds extend over even larger degree-wide ($\sim$50\,pc at 3\,kpc) scales.  Again, it is tempting to draw a connection between these scales and the similar sizes of GMCs and classic HII regions, 10--100\,pc.

We note an interesting consequence of this radiative transfer approach.  In Figure \ref{ratios6}, $\tau_{\rm 18}$ maximises near a value of 0.6; for any higher optical depth, all CO isotopologues would begin to saturate at $T_{\rm mb}$ = $T_{\rm ex}-T_{\rm bg}$.  Such clouds would have $\tau$ indistinguishable from infinity, whatever the $T_{\rm ex}$ (all of the $R_{\rm 18}$ curves converge to where both line ratios are 1).  Even so, $\tau_{\rm 18}$ = 0.6 corresponds approximately to $\tau_{\rm 12}\sim250$ for any reasonable values of the abundance ratios $R$.  This means that there are a large number of clouds in the ThrUMMS maps with very large column densities (e.g., near the top-right of the plots in Fig.\,\ref{ratios6}, with blue, cyan, or green ratios), some of which could masquerade as lower column densiy clouds if one were to convert from $I_{\rm 12}$ with a standard X-factor.

\subsection{Column Density Maps}\label{coldens}
A subject of much study in the literature is the conversion of observed molecular emission into estimates of molecular column density or mass, which are important for a number of questions in astrophysics, e.g., in star formation when measuring cloud stability against gravitational collapse, or across disk galaxies for studies of the Kennicutt-Schmidt relation, or for estimating the contribution of the molecular ISM to a galaxy's dynamics, etc.  \citet{dht01} summarised this topic and, when converting CO emission into mass, obtained a best average estimate for the X-factor of\vspace{2mm}
\begin{displaymath}
	\hspace{-55.5mm}X_{\rm CO} = N_{H_2}/I_{\rm CO}\vspace{-2mm}
\end{displaymath}
\begin{equation} 
	= 1.8\times10^{24}~{\rm H}_2\,{\rm molecules\,m}^{-2}/({\rm K\,km\,s}^{-1})
\end{equation}
in the solar neighbourhood, based on high-latitude data in the CfA survey.  But as \citet{dht01} point out, this value is expected to vary locally in the Galactic midplane (i.e., at low latitudes) due to various factors.

Having derived $\tau$ and $T_{\rm ex}$ at each voxel, we can use the ThrUMMS data to make a unique contribution to this exercise with a CO column density map, via
\begin{equation} 
	N_{\rm CO} = \frac{3h}{8\pi^3\mu^2}~\frac{Q(T_{\rm ex})e^{E_u/kT_{\rm ex}}}{J_u(1-e^{-h\nu/kT_{\rm ex}})}~\int\tau{\rm d}V  ,
\end{equation}
where $\mu$ is the CO dipole moment, $Q$ is the rotational partition function, $E_u$ and $J_u$ are the energy and quantum number of the upper level of the transition at frequency $\nu$, and the integral is over the velocity range of the emission line.  Here, instead of using a constant $X_{\rm CO}$ factor to convert the observed $I_{\rm CO}$ into a molecular gas mass, as is sometimes done in the literature despite \citet{dht01}'s caveat against doing so, we can use the ($\tau$,$T_{\rm ex}$) estimates above (\S\ref{tauratios}) to compute a more realistic CO column density.  This can then be compared with other estimates of molecular gas mass (e.g., SED fits to the Hi-GAL data) to derive CO abundance maps.

The computation of eq.\,(5) across a spatially-resolved map with $\sim$8$\times$10$^8$ independent ($l$,$b$,$V$) resolution elements, complete with a calculation of the partition function at each voxel and appropriately propagating uncertainties in each quantity, is a non-trivial exercise.  Moreover, in many locations the \ttco\ and especially \ceto\ can be quite weak, so there are S/N limitations in applying this method to our full-resolution data cubes.  Nevertheless, as described in \S\ref{tauratios} we have performed this computation on cubes that have been binned in velocity to 1\kms\ channels.  After so computing the $\tau$ and $T_{\rm ex}$ in each such voxel, we use eq.\,(5) to derive cubes of column density per channel.

We also describe here a simpler calculation, which gives a very good approximation to eq.\,(5) in cases where insufficient computational power is available.  The temperature-dependent terms in eq.\,(5) can be collected and written as a closed function of $T_{\rm ex}$, which we call the modified temperature function ${\mathfrak T}_{\rm mod}$:
\begin{equation} 
	{\mathfrak T}_{\rm mod} = \frac{Q(T_{\rm ex})e^{E_u/kT_{\rm ex}}}{J_u(1-e^{-h\nu/kT_{\rm ex}})}~.
\end{equation}
Over a typical range of $T_{\rm ex}$ = 5--100\,K, the following formula for ${\mathfrak T}_{\rm mod}$ gives deviations of $<$1\% from the exact expression (eq.\,6), without having to calculate the partition function $Q$ explicitly:
\begin{equation} 
	{\rm log}^2 ({\mathfrak T}_{\rm mod}) = 1 + {\rm log}^2 [0.053~T_{\rm ex}^{1.97}]~.
\end{equation}

Having computed the ${\mathfrak T}_{\rm mod}$ cube from the \tex\ cube using either eq.\,(6) or (7), we then obtain the column density cube from this and the $\tau$ cube, 
\begin{equation} 
	N_{\rm CO} = 5.57\times10^{18}\,{\rm m}^{-2}~{\mathfrak T}_{\rm mod}\int\tau{\rm d}V~,
\end{equation} 
where the integral over $V$ means per 1\kms\ channel.  We emphasise, however, that we have actually done the full calculation from eq.\,(5/6) in this study.

A moment map of the column density cube from eq.\,(5) is shown in Figure \ref{tautex} (bottom panel).  This now embodies a much larger ``conversion factor'' than in \citet{dht01}.  In the cool and opaque case, we might have typical values $\tau$\gapp100 and $T_{\rm ex}$$\sim$10\,K.  For a cloud with an effective line width $\sim$5\,\kms, $N_{\rm CO}\sim$ 4.1$\times$10$^{22}$\,{\rm m}$^{-2}$.  
In fact, the bottom panel of Figure \ref{tautex} shows that column densities computed in this way can reach a peak value some 5$\times$ higher than this, or $N_{\rm CO}\sim$ 2$\times$10$^{23}$\,{\rm m}$^{-2}$.  Assuming that the \tco\ abundance is 10$^{-4}$ (although see \S\ref{heavy}), this is equivalent to $N_{H_2}\sim$ 0.4--2$\times$10$^{27}$\,{\rm m}$^{-2}$, or a mass column $\Sigma\sim$ 760--3700\,M\solar\,pc$^{-2}$.  
Even the lower end of this range is close to the mass columns in parsec-scale ``dense clumps'' seen in previous surveys \cite[e.g.,][]{b11}, but is much higher than typical GMC mass columns.  We can convert the above cool, opaque example to an equivalent $X_{\rm CO}$$\sim$ 8--40$\times$10$^{24}$\,m$^{-2}$/(K\kms), and so realise that the Milky Way's molecular gas mass may need to be revised upwards by a significant amount, at least in some locations; see the discussion in \S\ref{heavy}.

Equally remarkable as the actual numbers are the structural revelations in Figure \ref{tautex}.  As has been seen in many previous studies, the integrated intensity iso-CO maps (Appendix, Fig.\,\ref{sector6}--\ref{sector1}) have a rather complex combination of structures.  The respective data cubes reveal that this complexity breaks down into individual regions or clouds at various \vlsr, which may be clumpy and compact, or filamentary, or amorphous on larger angular scales.  However, the cubes by themselves do not reveal any obvious systematic trend among these structures.

In contrast, the $\tau$, \tex, and $N_{\rm CO}$ cubes show systematically different structural characteristics.  As already noted in \S\ref{tauratios}, the $\tau$ cubes are clumpy and high-contrast, while the \tex\ cubes are amorphous and low-contrast.  In Figure \ref{tautex}, the $N_{\rm CO}$ panel (bottom) reveals the combination of the $\tau$ and \tex: a high-contrast and more filamentary medium than either of the above constituents.  One is naturally drawn to compare such images with the almost ubiquitous filamentary structures evident in the Hi-GAL maps.

\notetoeditor{}
\begin{figure*}[ht]
\centerline{\includegraphics[angle=0,scale=0.275]{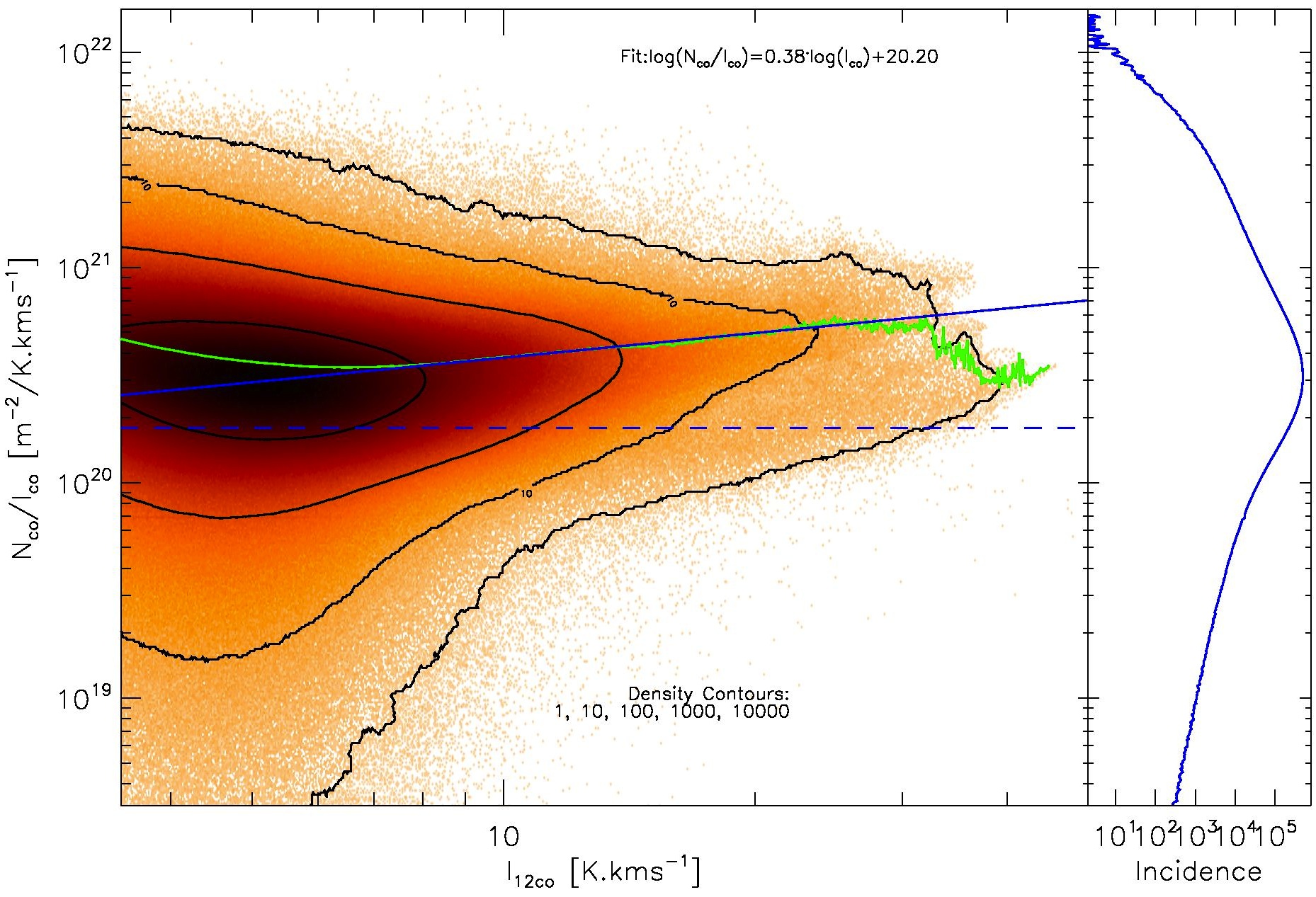}} 
\vspace{0mm}
\caption{
Voxel incidence of the ratio of computed column density from eq.\,(5) to the measured \tco\ integrated intensity per 1\kms\ binned channel, plotted as a function of the integrated intensity, across the 4Q.  The green line represents mean values of this ratio in narrow bins of $I_{\rm CO}$; the dashed blue line shows the expected $X_{\rm CO}$/$R_{\rm 12}$ ratio, or equivalently, the conversion of $I_{\rm CO}$ to $N_{\rm H_2}$ using eq.\,(4) with standard values of $X$ and $R$; while the sloping blue line is a weighted power-law fit to the modal trend, as labelled at the top (see text for further discussion).  The histogram to the right is effectively a distribution function for $X$/$R$: a column-weighted mean value is 5.1$\times$10$^{20}$\,CO molecules\,m$^{-2}$/(K\kms).
}
\label{NvsI}
\end{figure*}

\subsection{Global Implications}\label{heavy}

Based on the distribution of points in Figure \ref{ratios6} and the conversion embodied in eq.\,(5), we can in principle compute mass, density, column density, or any other distribution functions 
for the entire molecular cloud population of the 4Q, such as in Figure \ref{NvsI} (discussed next).  For many derived parameters (such as mass), we also need distances to individual clouds to obtain them from the column density computed here.  However, assignment of distances to all the structures we see in the cubes is far beyond the scope of this paper.

Despite this, our results have a number of significant implications for studies of molecular cloud populations in disk galaxies.  For example, a variety of observational measurements and chemical models indicate that the fractional CO abundance relative to molecular hydrogen typically\footnote{
This is an average across different environments throughout the Milky Way: lower values (10$^{-4.5}$) seem appropriate in dark clouds --- our cool, opaque case --- and higher values (10$^{-3.5}$) in luminous HII regions --- our warm, translucent case \cite[e.g., see][]{g14}.  
} lies in the range 10$^{-3.5}$ to 10$^{-4.5}$; a mean value of 10$^{-4}$ is often used in the literature.  As discussed in the last section, this means that we should be able to simply convert our computed $N_{\rm CO}$ to $N_{\rm H_2}$ by such a factor.  But we also have the standard conversion of $I_{\rm CO}$ to $N_{\rm H_2}$ using eq.\,(4).  For these two conversions to be consistent, the various quantities must be related.  To make these relationships clearer, we rewrite eq.\,(4) as
\begin{equation} 
	N_{\rm H_2} = X_{\rm CO} I_{\rm CO} = R_{\rm 12} N_{\rm CO}~,
\end{equation}
where we define the ratio $R_{\rm 12}$ from the actual CO abundance in the gas, $R_{\rm 12} \equiv N_{\rm H_2}$/$N_{\rm CO}$, or typically 10$^{+3.5\,{\rm to}\,+4.5}$ from above.  Note that this definition makes the gas abundance ratio $R_{\rm 12}$ different to the empirically determined $X_{\rm CO}$ conversion factor.  Then 
\begin{equation} 
	N_{\rm CO} / I_{\rm CO} = X_{\rm CO} / R_{\rm 12}~.
\end{equation}
With the standard values on the RHS of eq.\,(10), the ratio on the LHS should be a constant,
\begin{equation} 
	N_{\rm CO} / I_{\rm CO} = 1.8\times10^{20}~{\rm CO\,molecules\,m}^{-2}/({\rm K\,km\,s}^{-1}).
\end{equation}
However, when we plot eq.\,(11) vs.\,$I_{\rm CO}$ in Figure \ref{NvsI}, we see a marked variation in the actual ratio $X$/$R$ as shown in eq.\,(10), instead of the flat line that would be expected from eq.\,(11) with fixed values of $X$ and $R$ (shown as a dashed blue line in Fig.\,\ref{NvsI}).

Even at the lowest values of $I_{\rm CO}$ (\lapp\,6\,K\kms\ per 1\kms-wide channel) in Figure \ref{NvsI}, the modal fit (blue line) of $X$/$R$ already begins above the standard value (dashed line), while the mean values (green line) lie around 3--4$\times$10$^{20}$ CO molecules m$^{-2}$/(K\kms), or twice the standard value.  If one assumes $X_{\rm CO}$ is fixed to the standard value, this implies that $R_{\rm 12}$ = 5$\times$10$^3$, or half its standard value.  On the other hand, if one assumes $R_{\rm 12}$ is fixed to its standard value, this implies that $X_{\rm CO}$ = 3--4$\times$10$^{24}$ H$_2$ molecules m$^{-2}$/(K\kms), or twice its standard value.  As $I_{\rm CO}$ rises, the modal fit in $X$/$R$ also rises further to almost 10$^{21}$ CO molecules m$^{-2}$/(K\kms) at $I_{\rm CO}$ = 40\,K\kms\ per 1\kms-wide channel.  This means that at these brightness levels, either $X_{\rm CO}$ is 4$\times$ higher than standard, or $R_{\rm CO}$ is 4$\times$ lower than standard, or some combination of these two options applies.  At $I_{\rm CO}>$ 20\,\kms, the $X_{\rm CO}$/$R_{\rm 12}$ distribution actually bifurcates\footnote{
This bifurcation, which also appears in the left panel of Fig.\,\ref{ratios34}, appears to be entirely due to the two brightest GMC complexes in \tco\ in our maps, but these still illustrate how the distribution of points in Fig.\,\ref{NvsI} depends on the iso-CO ratios.  The upper branch comes from the G333 complex, which is bright in all three CO lines, and so has moderate optical depth and high column density.  The lower branch comes from the G351.3+0.6 complex, which is bright in \tco\ only, and so has lower optical depth and lower column density.}
 into two regimes, one with the ratio up to 2$\times$ higher than the standard value, and another up to 10$\times$ higher.  A weighted power-law fit to the upper trend (sloping blue line in the plot) gives
\begin{equation} 
	X/R \propto {I_{\rm CO}}^{0.38}
\end{equation}
across the measured range of $I_{\rm CO}$, or equivalently, 
\begin{equation} 
	N_{\rm CO} = 1.6 \times 10^{20}~{\rm m}^{-2}~I_{\rm CO}^{1.38}  .  
\end{equation}
This is a remarkable result for several reasons, but mainly because in retrospect, it should have been more widely recognised before now.  For example, the fact that eq.\,(13) is superlinear falls naturally out of our very simple radiative transfer assumptions, and theoretically was predicted on this basis \citep{kth07,ncs08}.  Also, because eq.\,(13) is superlinear, it is no longer as physically meaningful to talk about a single X-factor: the direct conversion from CO integrated intensity to hydrogen column density can be seen as an illusion that is correct only in the mean, and masks the very real effects of high optical depth in the \tco\ line.  Thus, the scale factor in eq.\,(13) has units of molecules m$^{-2}$/(K\kms)$^{1.38}$, and so dimensionally is not a ``conversion factor.''

Nevertheless, we can compute average quantities from our data, which may still be useful in non-spatially-resolved studies, such as extragalactic or high-redshift star-formation settings.  The equivalent median value of the X-factor for {\em all} our data is $X_{\rm CO}$ = 3.8$\times$10$^{24}$\,m$^{-2}$/(K\kms) or 2.1$\times$ the standard value, while a {\em column-weighted} median conversion is even larger than this, $X_{\rm CO}$ = 5.1$\times$10$^{24}$\,m$^{-2}$/(K\kms) or 2.8$\times$ the standard value.

However, eqs.\,(12--13) represent just the modal trend in Figure \ref{NvsI}.  It should be clear that the range of the variations in $X_{\rm CO}$/$R_{\rm 12}$ is far too large to be attributed to changes in just one of these parameters.  Even the 1$\sigma$ variation from the modal trend in Figure \ref{NvsI} is 0.3 in the log, or a factor of 2.  But the extrema in Figure \ref{NvsI} show $X$/$R$ ratios from about 30$\times$ larger to 100$\times$ smaller than standard.  This represents $>$3 orders of magnitude empirical uncertainty in the conversion of $I_{\rm CO}$ to $N_{\rm H_2}$.  This variation exceeds any uncertainty in the assumptions we have used in this analysis, such as the assumption of LTE at a single \tex, or values of the ratios $R$, or even of the constancy of $X_{\rm CO}$ in the face of the metallicity gradient in the Galactic disk \citep{hd15}.

Such uncertainties aside, in the mean it is likely that, as $I_{\rm CO}$ rises from 4 to 40\,K\kms\ in each 1\kms-wide channel, $X_{\rm CO}$ is very likely rising {\em and that} $R_{\rm 12}$ is probably falling (meaning that the gas phase abundance of \tco\ relative to H$_2$ is probably rising).  Therefore, the conversion of $I_{\rm CO}$ to an equivalent mass surface density is not as simple as has often been assumed.  While we cannot yet derive the exact corrections to cloud masses, and the Milky Way's total molecular mass, without clouds' individual distances, this result has obvious implications for studies of the molecular mass distribution in the Milky Way, star formation laws in disk galaxies such as the Schmidt-Kennicutt relations, and other topics \citep{L08,L15}.  At the very least, it implies that such relations must be recalibrated for the highest optical depth clouds and regions, which in our data are widely distributed across the Galactic Plane, and seem to have not yet been adequately taken into account.

\section{Conclusions}\label{concl}

We have presented information about, and first data releases and science results from, a comprehensive new mm-wave multi-spectral-line survey of the southern Milky Way, ThrUMMS.  Essential features of this survey are its public access, speed, coverage (120 deg$^2$), high angular (arcminute) and physical (parsec) resolution, and simultaneous mapping of four important molecular tracers, \tco, \ttco, \ceto, and CN.  This combination of attributes makes ThrUMMS unique.  We have also presented first science results on the global properties of the Galactic population of molecular clouds in the Fourth Quadrant.  These include the discovery of widespread and large variations in the line ratios of the three main CO isotopologues with position and velocity, and the implication that some of these variations arise in a large population of clouds with very high column density.  We derive a new calibration of the CO-to-H$_2$ conversion, embodied by our eq.\,(13) and replacing the concept of a fixed $X_{\rm CO}$.  With an eventual catalogue of cloud distances, this will drive related corrections to star-formation and other laws.  We anticipate a wide range of further science applications to spring from this unique data set for many years into the future.

\acknowledgments

We thank the staff members of the ATNF for their support of the Mopra telescope and our observations.  PJB thanks Tom Dame for challenging him to find a way to ``do it faster,'' and acknowledges support from NASA/JPL contract RSA-1464327, NSF grants AST-0903672 (to C. Telesco) and AST-1312597, and the University of Florida.  AKH acknowledges support from  HST grant GO-12275 and NASA-ADP grant NNX11AD18G to B. Wakker.  This work makes extensive use of IDL libraries developed by NASA (http://idlastro.gsfc.nasa.gov), David Fanning (http://www.idlcoyote.com), and James Davenport.

{\it Facilities:} \facility{Mopra (MOPS)}

\clearpage



\appendix
\section{ThrUMMS Integrated Intensity Images}

We present here all integrated intensity images across the 4Q survey area, broken up into one 10\degree$\times$2\degree\ sector per printed page.  Each sector is shown both as a combined pseudo-colour image with red = \tco, green = \ttco, and blue = \ceto, and beneath this, four separate images for each of the four species mapped, as labelled.  A single \tco\ contour at 28.94\,K\kms\ $\approx$ 20$\sigma$ is shown in each of the four species' panels, labelled in the \tco\ panel.  Note that CN (yellow) is not included in the pseudo-colour images.

\notetoeditor{}
\begin{figure*}[h]

\vspace{9mm}
\includegraphics[angle=0,scale=0.35]{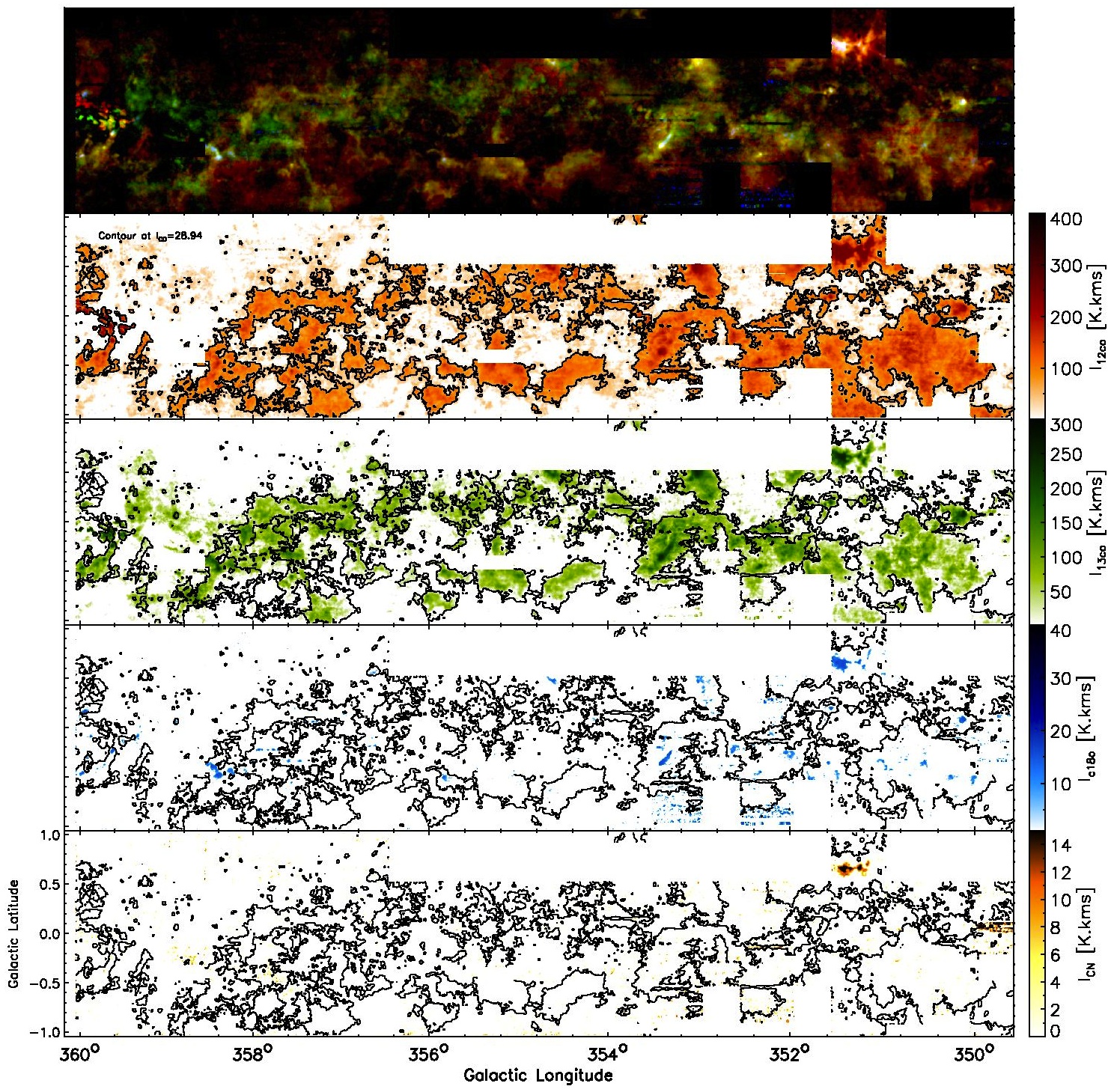}
\caption{ThrUMMS DR3 integrated intensity images for 360\degree$>l>$350\degree.
}
\label{sector6}
\end{figure*}

\notetoeditor{}
\begin{figure*}[h]

\vspace{20mm}
\includegraphics[angle=0,scale=0.35]{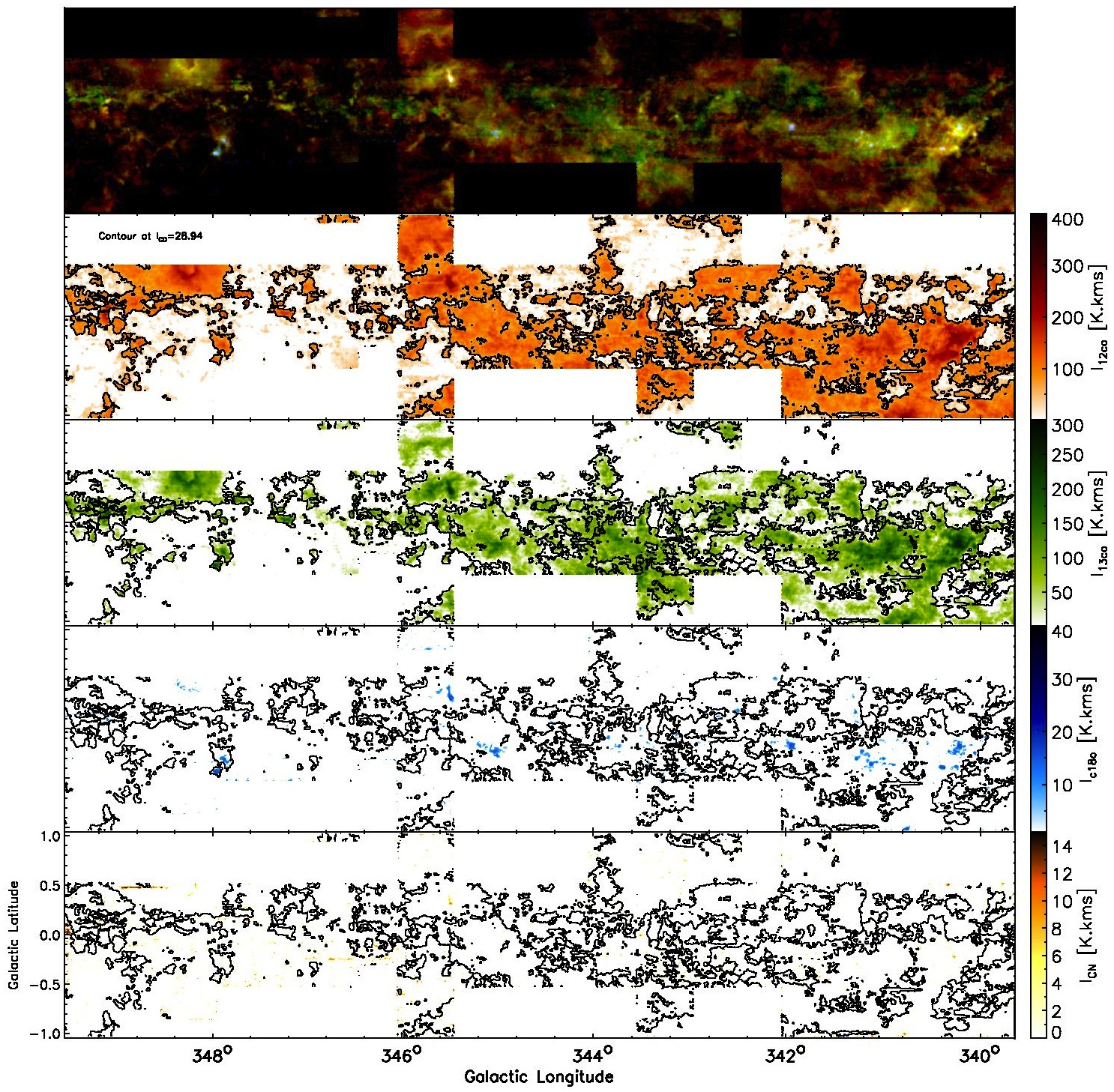}
\caption{ThrUMMS DR3 integrated intensity images for 350\degree$>l>$340\degree.
}
\label{sector5}
\end{figure*}

\notetoeditor{}
\begin{figure*}[h]

\vspace{20mm}
\includegraphics[angle=0,scale=0.35]{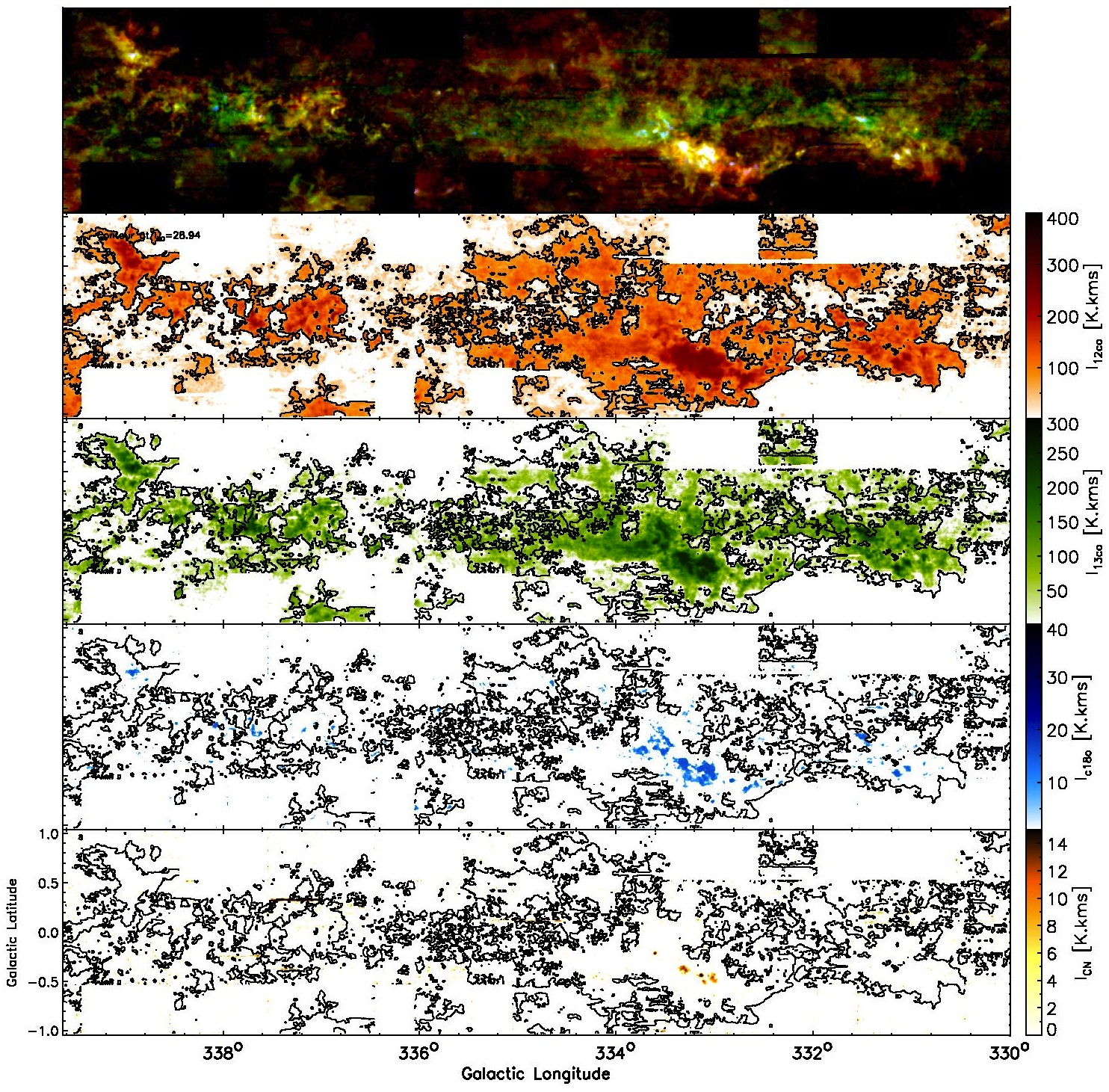}
\caption{ThrUMMS DR3 integrated intensity images for 340\degree$>l>$330\degree.
}
\label{sector4}
\end{figure*}

\notetoeditor{}
\begin{figure*}[h]

\vspace{20mm}
\includegraphics[angle=0,scale=0.35]{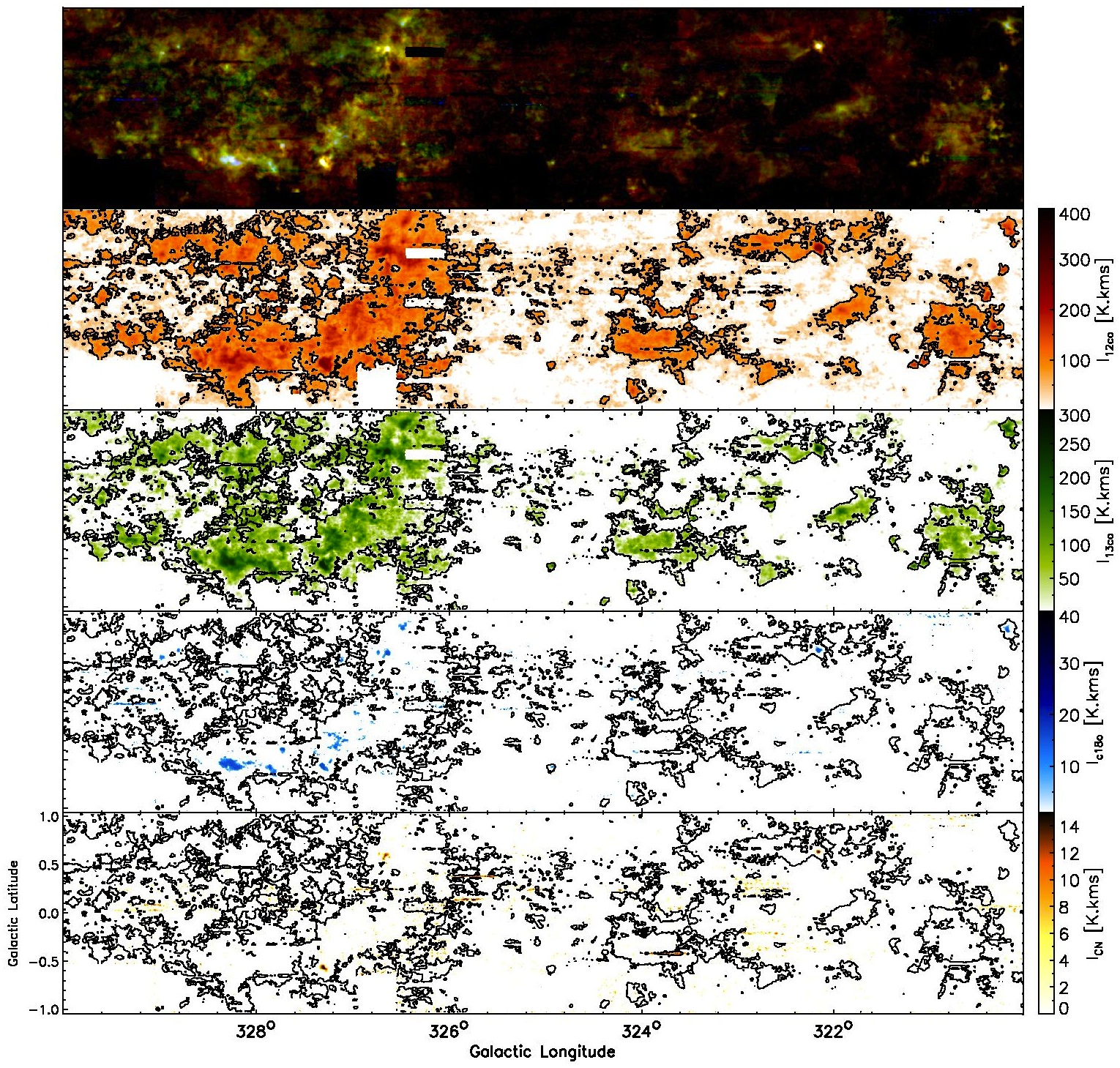}
\caption{ThrUMMS DR3 integrated intensity images for 330\degree$>l>$320\degree.
}
\label{sector3}
\end{figure*}

\notetoeditor{}
\begin{figure*}[h]

\vspace{20mm}
\includegraphics[angle=0,scale=0.35]{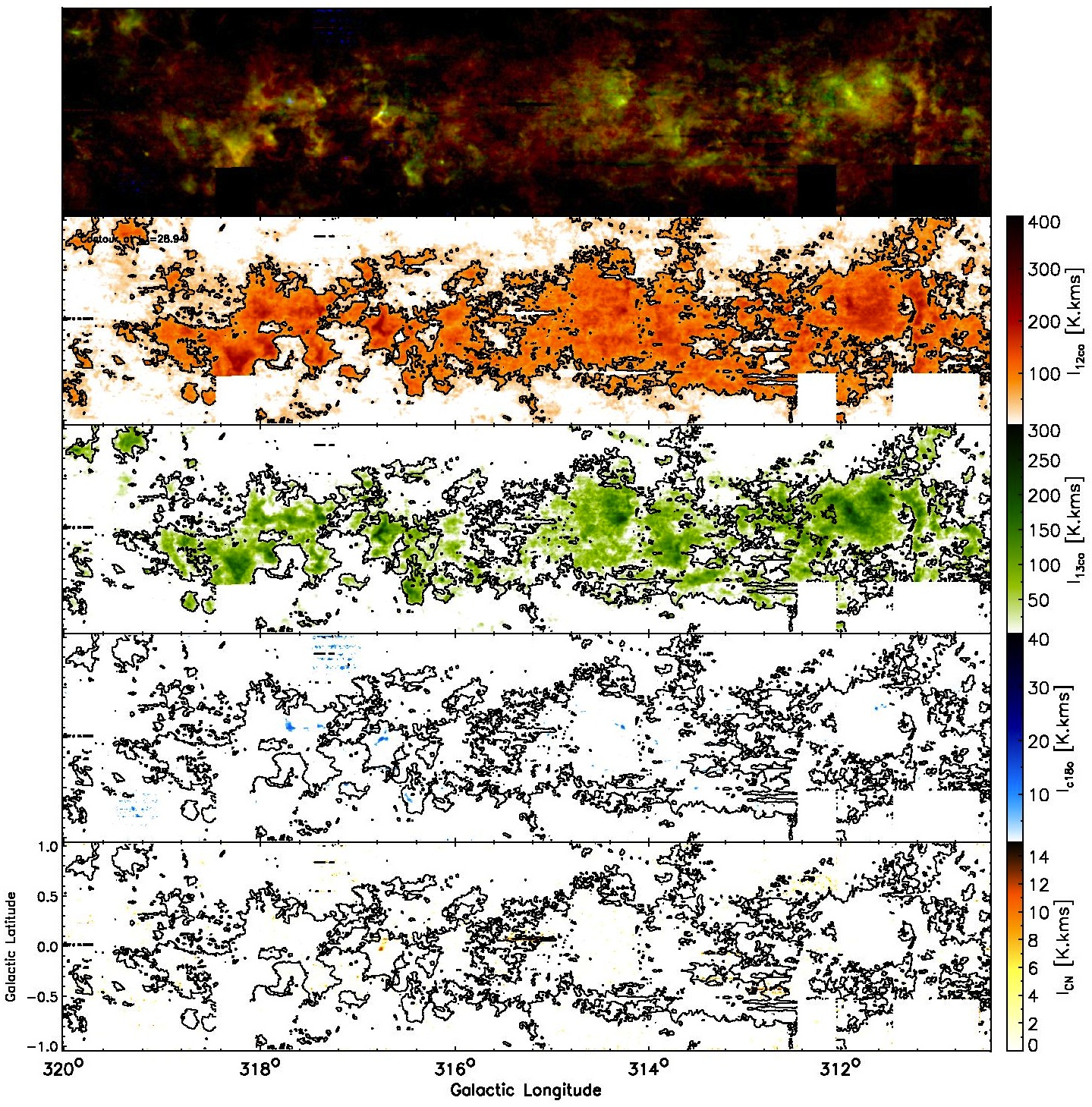}
\caption{ThrUMMS DR3 integrated intensity images for 320\degree$>l>$310\degree.
}
\label{sector2}
\end{figure*}

\notetoeditor{}
\begin{figure*}[h]

\vspace{20mm}
\includegraphics[angle=0,scale=0.35]{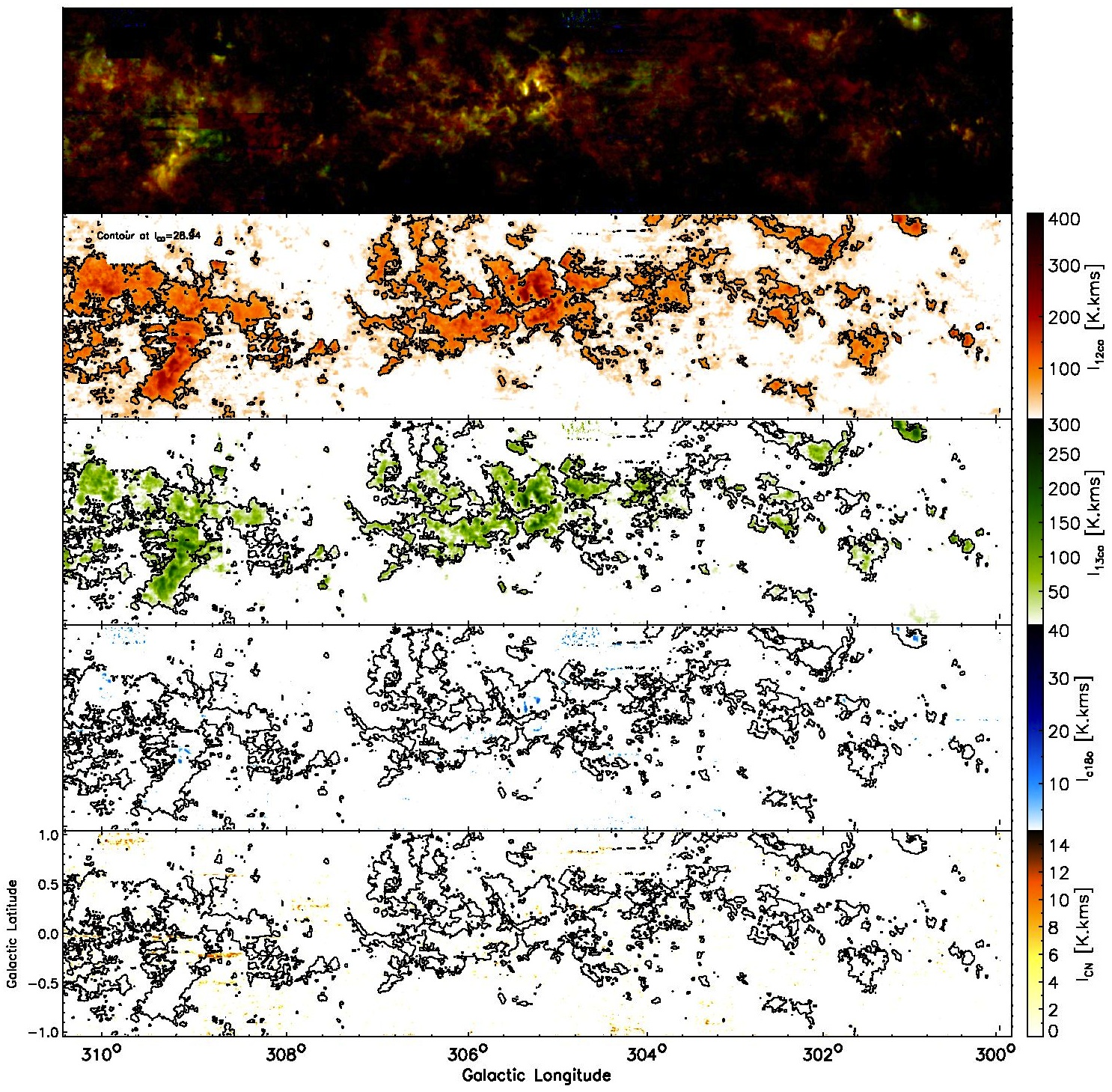}
\caption{ThrUMMS DR3 integrated intensity images for 310\degree$>l>$300\degree.
}
\label{sector1}
\end{figure*}

\end{document}